\documentclass[11pt,a4paper]{article}
\usepackage{imakeidx}
\makeindex[columns=2, title=Alphabetical Index, intoc]
\usepackage{jheppub} % for details on the use of the package, please see the JINST-author-manual
\usepackage[pagewise]{lineno}
\usepackage{amsmath}
\usepackage{dsfont} %mathds font
\usepackage{nicefrac}
\usepackage[dvipsnames]{xcolor}
\usepackage{braket}
\usepackage{enumitem}
\usepackage{mathtools}
\usepackage{cleveref}
\usepackage{bbold}
\usepackage{physics}

\usepackage{multirow}
\usepackage{caption}
\usepackage{makecell}
\usepackage{subcaption}

\usepackage{graphicx}
\usepackage{tikz-feynman}
\usetikzlibrary{calc}
\tikzfeynmanset{compat=1.1.0,warn luatex=false}
\usepackage{mathdots}
\usepackage{textgreek}

\usepackage{rotating}
\usepackage{cancel}
\usepackage{slashed}
\usepackage{soul}

\usepackage{longtable} 

\newcommand{\be}{\begin{equation}}
\newcommand{\ee}{\end{equation}}
\makeatletter
\newcommand\ps@indexpagestyle{
 \renewcommand\@oddfoot{\hfill-- \thepage\ --\hfill}
 \renewcommand\@oddhead{}
}

\makeatother
\indexsetup{firstpagestyle=indexpagestyle}

\def\KITITP{Institute for Theoretical Physics, KIT, Wolfgang-Gaede-Straße 1, 76131, Karlsruhe, Germany}
\def\PITT{Pittsburgh Particle-physics Astro-physics \& Cosmology Center (PITT-PACC), Department of
Physics $\&$ Astronomy, University of Pittsburgh, Pittsburgh, USA}
\def\KITA{Institute for Theoretical Particle Physics, KIT, Wolfgang-Gaede-Straße 1, 76131, Karlsruhe, Germany}

\preprint{
\begin{flushright} 
KA-TP-29-2026 \\
TTP26-041 \\
P3H-26-088
\end{flushright}
}
\title{
Absence of linear power corrections in
heavy-quark pair production in a
non-Abelian gauge theory}

\author[a,b]{Duarte Fontes,}
\author[c]{Dennis Horstmann,}
\author[c]{Kirill Melnikov,}
\author[c]{Davide Maria Tagliabue}

\emailAdd{duarte.fontes@pitt.edu}
\emailAdd{dennis.horstmann@kit.edu}
\emailAdd{kirill.melnikov@kit.edu}
\emailAdd{davide.tagliabue@kit.edu}

\affiliation[a]{\KITITP}
\affiliation[b]{\PITT}
\affiliation[c]{\KITA}

\abstract{We study the infrared sensitivity of heavy-quark pair production by a color-neutral vector current in perturbation  theory. We work in a spontaneously broken ${\rm SU}(2)$ gauge theory where, by tuning parameters to make the gauge-boson mass $\mg$ small, we use it as a probe of infrared physics. We find that, through next-to-next-to-leading order, terms linear in $\mg$ cancel in the differential cross section, provided that it is expressed through a short-distance heavy-quark mass.
Interestingly, we establish this cancellation without evaluating any integrals explicitly, by relating virtual integrals to real-emission ones through loop-energy residues, with a suitable redefinition of the real-emission momenta.
Our result is the first strong indication that the cancellation of non-perturbative linear power corrections in the corresponding QCD process extends to the non-Abelian sector. It is also a concrete non-Abelian example of the extension of the Kinoshita-Lee-Nauenberg theorem to linear infrared terms for sufficiently 
inclusive observables. 
}

\keywords{QCD corrections, IR sensitivity, hadronic colliders, NNLO calculations}

\usepackage{physics}
\usepackage{graphicx}

\usepackage{bbding}
\usepackage{mathtools}
\usepackage{booktabs}
\usepackage{lscape}
\usepackage{longtable,setspace} 
\usepackage{url}
\usepackage{bbold}
\usepackage{bm}
\usepackage{csquotes}

\usepackage{xspace}
\usepackage{makecell}
\usepackage{xifthen}

\usepackage{pgfplots}
\usepgfplotslibrary{colorbrewer}
\pgfplotsset{compat=newest}
\pgfplotsset{colormap/violet}

\usepackage{caption}
\usepackage{subcaption}
\usepackage{feynmp-auto}
\usepackage{color}
\definecolor{mycolor}{RGB}{165,111,29}
\usepackage{pdfpages}
\usepackage{needspace}
\usepackage[normalem]{ulem}

\usepackage[most]{tcolorbox}

\definecolor{darkbrown}{RGB}{150,75,10}
\definecolor{dmtgreen}{RGB}{0,120,80}

\usepackage{dowith}

\newcommand{\UpperCase}[1]{
  \expandafter\newcommand\csname bb#1\endcsname{{\mathbb{#1}}}
  \expandafter\newcommand\csname cal#1\endcsname{{\mathcal{#1}}}   
  \expandafter\newcommand\csname rm#1\endcsname{{\mathrm{#1}}}
  \expandafter\newcommand\csname bf#1\endcsname{{\mathbf{#1}}}
  \expandafter\newcommand\csname bold#1\endcsname{{\boldsymbol{#1}}}
  \expandafter\newcommand\csname hat#1\endcsname{\hat{#1}}
  \expandafter\newcommand\csname tilde#1\endcsname{\widetilde{#1}}
  \expandafter\newcommand\csname bar#1\endcsname{\overline{#1}}
  \expandafter\newcommand\csname frak#1\endcsname{\mathfrak{#1}}
  }
\DoWith\UpperCase ABCDEFGHILJKMNOPQRSTUVWXYZ\StopDoing

\newcommand{\LowerCase}[1]{
  \expandafter\newcommand\csname rm#1\endcsname{{\mathrm{#1}}} 
  \expandafter\newcommand\csname bf#1\endcsname{{\mathbf{#1}}} 
  \expandafter\newcommand\csname bold#1\endcsname{{\boldsymbol{#1}}}
  \expandafter\newcommand\csname hat#1\endcsname{\hat{#1}}
  \expandafter\newcommand\csname tilde#1\endcsname{\tilde{#1}}
  \expandafter\newcommand\csname bar#1\endcsname{\bar{#1}}
  \expandafter\newcommand\csname frak#1\endcsname{\mathfrak{#1}}
  \expandafter\newcommand\csname vec#1\endcsname{\vec{#1}}
  }
\DoWith\LowerCase abcdefghijklmnopqrstuvwxyz\StopDoing

\newcommand{\MeasureUpperCase}[1]{
  \expandafter\newcommand\csname d#1\endcsname{\mathrm{d}#1}
  \expandafter\newcommand\csname dvec#1\endcsname{\mathrm{d}\vec{#1}}
  }
\DoWith\MeasureUpperCase ABCDEFGHILJKMNOPQRSTUVWXYZ\StopDoing

\newcommand{\MeasureLowerCase}[1]{
  \expandafter\newcommand\csname d#1\endcsname{\mathrm{d}#1}
  \expandafter\newcommand\csname dvec#1\endcsname{\mathrm{d}\vec{#1}}
  }
\DoWith\MeasureLowerCase abcegiklmnrstuwxyz\StopDoing

\newcommand{\lint}{\big\langle}
\newcommand{\rint}{\big\rangle}

\newcommand{\autoL}{\left\langle}
\newcommand{\autoR}{\right\rangle}

\newcommand{\SU}[1]{{\rm SU}(#1)}

\newcommand{\IRR}[1]{I_{\rm RR}^{(#1)}}
\newcommand{\IRV}[1]{I_{\rm RV}^{(#1)}}
\newcommand{\IVV}[1]{I_{\rm VV}^{(#1)}}
\newcommand{\IR}[1]{I_{\rm R}^{(#1)}}
\newcommand{\IV}[1]{I_{\rm V}^{(#1)}}

\newcommand{\mg}{m_g}
\newcommand{\mh}{m_H}

\newcommand{\mt}{m_t}
\newcommand{\mtt}{\widetilde{m}_t}

\newcommand{\rbchi}{\chi}
\newcommand{\pt}{p_t}
\newcommand{\ptb}{p_{\bart}}
\newcommand{\qt}{q_t}
\newcommand{\qtb}{q_{\bart}}
\newcommand{\om}{\omega}

\newcommand{\Fm}{u}
\newcommand{\Sm}{r}

\newcommand{\iZero}{i0}

\newcommand{\barpsi}{\overline{\psi}}

\newcommand{\ghost}{{\rm Ghost}}
\newcommand{\GF}{{\rm GF}}

\newcommand{\eq}{Eq.~}

\newcommand{\ep}{\epsilon}

\newcommand{\mydots}{...}

\newcommand{\sign}{{\rm sign}}

\newcommand{\rhs}{right-hand side}

\newcommand{\Sec}{Section~}

\newcommand{\TR}{T_\rmR}
\newcommand{\nf}{{n_\rmf}}

\newcommand{\Ca}{C_\rmA}
\newcommand{\Cf}{C_\rmF}

\newcommand{\as}{\alpha_{\rms}}

\newcommand{\gs}{g_{\rms}}

\newcommand{\psit}{\psi_t}

\usepackage[normalem]{ulem}

\usepackage{dashrule}

\newcommand{\bs}{\begin{subequations}}
\newcommand{\es}{\end{subequations}}

\definecolor{dgreen}{rgb}{0,0.2,0} 
\definecolor{mille}{rgb}{0.67,0.608,0.388} 
\definecolor{sbrown}{RGB}{139,69,19} 
\definecolor{myblueold}{RGB}{65,105,225}
\def\bd{\begin{dmath}} 
\def\ed{\end{dmath}}

\newcounter{DFQ}

\allowdisplaybreaks % split long formulas in two pages
\begin{document}

\maketitle
\flushbottom
% \linenumbers

%%%%%%%%%%%%%%%%%%%%%%%%%%%%%%%%%%%%%
%%% INPUT FIGURE and TABLES 
%%%%%%%%%%%%%%%%%%%%%%%%%%%%%%%%%%%%%
\newcommand{\TopPairTree}[1][1]{%
  \scalebox{#1}{%
    \begin{tikzpicture}[baseline={(current bounding box.center)}]
      \begin{feynman}
        \vertex (g)  at (-0.7,0);
        \vertex (v)  at ( 0,0);
        \vertex (t)  at ( 1.95, 1.15);
        \vertex (tb) at ( 1.95,-1.15);
        \diagram*{
          (g) -- [photon] (v),
          (v) -- [fermion,
                  momentum={[arrow shorten=0.35] $p_t$}] (t),
          (v) -- [anti fermion,
                  momentum'={[arrow shorten=0.35] $p_{\bar t}$}] (tb),
        };
      \end{feynman}
    \end{tikzpicture}%
  }%
}

\newcommand{\TopPairVirtual}[1][1]{%
  \scalebox{#1}{%
    \begin{tikzpicture}[baseline={(current bounding box.center)}]
      \begin{feynman}
        \vertex (g)  at (-0.7,0);
        \vertex (v)  at ( 0,0);
        \vertex (a)  at ( 0.95, 0.56);
        \vertex (b)  at ( 0.95,-0.56);
        \vertex (t)  at ( 1.95, 1.15);
        \vertex (tb) at ( 1.95,-1.15);
        \diagram*{
          (g) -- [photon] (v),
          (v) -- [fermion] (a)
              -- [fermion,
                  momentum={[arrow shorten=0.2075] $p_t$}] (t),
          (v) -- [anti fermion] (b)
              -- [anti fermion,
                  momentum'={[arrow shorten=0.2075] $p_{\bar t}$}] (tb),
          (a) -- [gluon, red] (b),
        };
      \end{feynman}
    \end{tikzpicture}%
  }%
}

\newcommand{\TopPairReal}[1][1]{%
  \scalebox{#1}{%
    \begin{tikzpicture}[baseline={(current bounding box.center)}]
      \begin{feynman}
        \vertex (g)  at (-0.7,0);
        \vertex (v)  at ( 0,0);
        \vertex (a)  at ( 0.90, 0.53);
        \vertex (b)  at ( 0.95,-0.56);
        \vertex (t)  at ( 2.05, 1.21);
        \vertex (tb) at ( 2.05,-1.21);
        % The emitted gluon is almost horizontal and points slightly
        % towards the anti-top direction, keeping k separate from q_t.
        \vertex (k)  at ( 2.20, 0.25);
        \diagram*{
          (g) -- [photon] (v),
          (v) -- [fermion] (a)
              -- [fermion,
                  momentum={[arrow shorten=0.2458] $q_t$}] (t),
          (v) -- [anti fermion,
                  momentum'={[arrow shorten=0.3573] $q_{\bar t}$}] (tb),
          (a) -- [gluon, red,
                  momentum'={[arrow shorten=0.2446]
                              \color{black}$k$}] (k),
        };
      \end{feynman}
    \end{tikzpicture}%
  }%
}
% To use the figure in the article, copy this definition
% and call \FigIVVone, \FigIRVone, etc. The tikz-feynman and graphicx packages are required.
% The optional argument scales the entire figure:
% \FigIVVone      = original size
% \FigIVVone[0.7] = 70% of the original size
% \FigIVVone[1.3] = 130% of the original size

%----------------------------------------
% FIGURE IVV1
%----------------------------------------
\newcommand{\FigIVVone}[1][1]{%
    \scalebox{#1}{%
    \begin{tikzpicture}[baseline={([yshift=0.3mm]c.base)}]
    \begin{feynman}
        % Same spiral dimensions as in IRR1 and IRV1.
        \tikzfeynmanset{gluon/.style={
            draw=none,
            decoration={mirror},
            postaction={draw,line width=0.65pt,decorate,decoration={
                coil,aspect=0.75,
                amplitude=3.6pt,segment length=2.16mm
            }}
        }}

        % Coordinates in cm: center, side vertices, and external lines.
        \vertex (c) at (0,0);
        \vertex (L) at (-2.8,0);
        \vertex (R) at ( 2.8,0);

        % External photons, each 0.6cm long.
        \vertex (i) at (-3.4,0);
        \vertex (f) at ( 3.4,0);

        % Open diamond endpoints on either side of the central line.
        \vertex (TL) at (-0.3, 2.5);
        \vertex (BL) at (-0.3,-2.5);
        \vertex (TR) at ( 0.3, 2.5);
        \vertex (BR) at ( 0.3,-2.5);

        % Three gluons meet at J, entirely within the left half.
        \vertex (J) at (-0.93, 0);
        % One attachment on the upper side and two on the lower side.
        \vertex (A) at (-1.32, 1.48);
        \vertex (C) at (-2.17,-0.63);
        \vertex (D) at (-1.0,-1.8);

        % The asterisk keeps the vertices at the positions specified above.
        % For black internal gluons, remove the three red options.
        % Diamond sides: 1.5pt; photons and gluons: 0.4pt.
        \tikzset{external line/.style={line width=1.5pt}}
        \diagram*{
            (i) -- [photon] (L);
            (R) -- [photon] (f);
            (TL) -- [plain, external line] (L) -- [plain, external line] (BL);
            (TR) -- [plain, external line] (R) -- [plain, external line] (BR);
            (J) -- [gluon, red] (A);
            (C) -- [gluon, red, decoration={mirror=true}] (J);
            (D) -- [gluon, red, decoration={mirror=true}] (J);
        };

        % Central line with two short horizontal segments.
        \draw [plain] (-0.38,-2.85) -- (0,-2.85)
                      -- (0,2.85) -- (0.38,2.85);
    \end{feynman}
    \end{tikzpicture}%
    }%
}

%----------------------------------------
% FIGURE IRV1
%----------------------------------------
% The optional argument scales the entire figure: \FigIRVone[0.45].
\newcommand{\FigIRVone}[1][1]{%
    \scalebox{#1}{%
    \begin{tikzpicture}[baseline={([yshift=0.3mm]c.base)}]
    \begin{feynman}
        % Style A: same spiral dimensions and line width as the other figures.
        \tikzfeynmanset{gluon/.style={
            draw=none,
            decoration={mirror},
            postaction={draw,line width=0.65pt,decorate,decoration={
                coil,aspect=0.75,
                amplitude=3.6pt,segment length=2.16mm
            }}
        }}

        % Same frame and external photons, with coordinates in cm.
        \vertex (c) at (0,0);
        \vertex (L) at (-2.8,0);
        \vertex (R) at ( 2.8,0);
        \vertex (i) at (-3.4,0);
        \vertex (f) at ( 3.4,0);
        \vertex (TL) at (-0.3, 2.5);
        \vertex (BL) at (-0.3,-2.5);
        \vertex (TR) at ( 0.3, 2.5);
        \vertex (BR) at ( 0.3,-2.5);

        % Two gluons connect the lower left side to the common vertex J.
        \vertex (C) at (-2.15,-0.65);
        \vertex (D) at (-1.10,-1.70);
        \vertex (J) at (-0.93,0);

        % The third gluon reaches the upper right side.
        \vertex (B) at (1.20,1.60);
        % U and E lie on J--B, leaving a gap around the central line.
        \vertex (U) at ($(J)!0.352!(B)$);
        \vertex (E) at ($(J)!0.521!(B)$);

        % Diamond sides: 1.5pt; photons and gluons: 0.4pt.
        % For black internal gluons, remove the four red options.
        \tikzset{external line/.style={line width=1.5pt}}
        \diagram*{
            (i) -- [photon] (L);
            (R) -- [photon] (f);
            (TL) -- [plain, external line] (L) -- [plain, external line] (BL);
            (TR) -- [plain, external line] (R) -- [plain, external line] (BR);
            (C) -- [gluon, red, decoration={mirror=true}] (J);
            (D) -- [gluon, red, decoration={mirror=true}] (J);
            (U) -- [gluon, red, decoration={mirror=true}] (J);
            (B) -- [gluon, red, decoration={mirror=true}] (E);
        };

        % Same central line with two short horizontal segments.
        \draw [plain] (-0.38,-2.85) -- (0,-2.85)
                      -- (0,2.85) -- (0.38,2.85);
    \end{feynman}
    \end{tikzpicture}%
    }%
}

%----------------------------------------
% FIGURE IRV2
%----------------------------------------
\newcommand{\FigIRVtwo}[1][1]{%
    \scalebox{#1}{%
    \begin{tikzpicture}[baseline={([yshift=0.3mm]c.base)}]
    \begin{feynman}
        % Same spiral dimensions as in IRR1.
        \tikzfeynmanset{gluon/.style={
            draw=none,
            decoration={mirror},
            postaction={draw,line width=0.65pt,decorate,decoration={
                coil,aspect=0.75,
                amplitude=3.6pt,segment length=2.16mm
            }}
        }}

        % Coordinates in cm: center, side vertices, and external lines.
        \vertex (c) at (0,0);
        \vertex (L) at (-2.8,0);
        \vertex (R) at ( 2.8,0);

        % External photons, each 0.6cm long.
        \vertex (i) at (-3.4,0);
        \vertex (f) at ( 3.4,0);

        % Open diamond endpoints on either side of the central line.
        \vertex (TL) at (-0.3, 2.5);
        \vertex (BL) at (-0.3,-2.5);
        \vertex (TR) at ( 0.3, 2.5);
        \vertex (BR) at ( 0.3,-2.5);

        % Two gluons meet at J from the upper and lower left sides.
        \vertex (A) at (-1.32, 1.48);
        \vertex (C) at (-2.02,-0.78);
        \vertex (J) at (-0.93, 0);

        % The third gluon continues towards the lower right side.
        \vertex (B) at (1.56,-1.24);

        % U and E lie on the straight line J--B.
        % These fractions leave a gap around the central line.
        \vertex (U) at ($(J)!0.30!(B)$);
        \vertex (E) at ($(J)!0.45!(B)$);

        % Styles: plain, photon (wavy), and gluon (spiral).
        % The asterisk keeps the vertices at the positions specified above.
        % For black internal gluons, remove the four red options.
        % The pair J--U / E--B shares a common axis.
        % post length leaves a 2.5mm straight segment near J.
        % Diamond sides: 1.5pt; photons and gluons: 0.4pt.
        \tikzset{external line/.style={line width=1.5pt}}
        \diagram*{
            (i) -- [photon] (L);
            (R) -- [photon] (f);
            (TL) -- [plain, external line] (L) -- [plain, external line] (BL);
            (TR) -- [plain, external line] (R) -- [plain, external line] (BR);
            (J) -- [gluon, red] (A);
            (C) -- [gluon, red, decoration={mirror=true}] (J);
            (U) -- [gluon, red,
                decoration={mirror=true,post length=1mm}] (J);
            (B) -- [gluon, red, decoration={mirror=true}] (E);
        };

        % Central line with two short horizontal segments.
        \draw [plain] (-0.38,-2.85) -- (0,-2.85)
                      -- (0,2.85) -- (0.38,2.85);
    \end{feynman}
    \end{tikzpicture}%
    }%
}

%----------------------------------------
% FIGURE IRR1
%----------------------------------------
% The optional argument scales the entire figure: \FigIRRone[0.45].
\newcommand{\FigIRRone}[1][1]{%
    \scalebox{#1}{%
    \begin{tikzpicture}[baseline={([yshift=0.3mm]c.base)}]
    \begin{feynman}
        % Style A: same spiral dimensions and line width as the other figures.
        \tikzfeynmanset{gluon/.style={
            draw=none,
            decoration={mirror},
            postaction={draw,line width=0.65pt,decorate,decoration={
                coil,aspect=0.75,
                amplitude=3.6pt,segment length=2.16mm
            }}
        }}

        % Same frame and external photons, with coordinates in cm.
        \vertex (c) at (0,0);
        \vertex (L) at (-2.8,0);
        \vertex (R) at ( 2.8,0);
        \vertex (i) at (-3.4,0);
        \vertex (f) at ( 3.4,0);
        \vertex (TL) at (-0.3, 2.5);
        \vertex (BL) at (-0.3,-2.5);
        \vertex (TR) at ( 0.3, 2.5);
        \vertex (BR) at ( 0.3,-2.5);

        % One gluon connects the upper left side to the common vertex J.
        \vertex (A) at (-1.63,1.17);
        \vertex (J) at (-0.93,0);

        % Two branches reach the upper and lower right sides.
        \vertex (B) at (1.50, 1.30);
        \vertex (K) at (1.20,-1.60);

        % Upper branch: U and E lie on J--B.
        \vertex (U) at ($(J)!0.309!(B)$);
        \vertex (E) at ($(J)!0.457!(B)$);

        % Lower branch: D and F lie on J--K.
        % Both branches leave a gap around the central line.
        \vertex (D) at ($(J)!0.352!(K)$);
        \vertex (F) at ($(J)!0.521!(K)$);

        % Diamond sides: 1.5pt; photons and gluons: 0.4pt.
        % For black internal gluons, remove the five red options.
        % Each pair of fragments shares the same straight axis.
        % The upper and lower branches have opposite spiral directions.
        \tikzset{external line/.style={line width=1.5pt}}
        \diagram*{
            (i) -- [photon] (L);
            (R) -- [photon] (f);
            (TL) -- [plain, external line] (L) -- [plain, external line] (BL);
            (TR) -- [plain, external line] (R) -- [plain, external line] (BR);
            (J) -- [gluon, red] (A);
            (U) -- [gluon, red, decoration={mirror=true}] (J);
            (D) -- [gluon, red, decoration={mirror=false}] (J);
            (B) -- [gluon, red, decoration={mirror=true}] (E);
            (K) -- [gluon, red, decoration={mirror=false}] (F);
        };

        % Same central line with two short horizontal segments.
        \draw [plain] (-0.38,-2.85) -- (0,-2.85)
                      -- (0,2.85) -- (0.38,2.85);
    \end{feynman}
    \end{tikzpicture}%
    }%
}

%----------------------------------------
% FIGURE IRR2
%----------------------------------------
\newcommand{\FigIRRtwo}[1][1]{%
    \scalebox{#1}{%
    \begin{tikzpicture}[baseline={([yshift=0.3mm]c.base)}]
    \begin{feynman}
        % Spiral dimensions in the reference figure.
        \tikzfeynmanset{gluon/.style={
            draw=none,
            decoration={mirror},
            postaction={draw,line width=0.65pt,decorate,decoration={
                coil,aspect=0.75,
                amplitude=3.6pt,segment length=2.16mm
            }}
        }}
        % Coordinates in cm: center, side vertices, and external lines.
        \vertex (c) at (0,0);
        \vertex (L) at (-2.8,0);
        \vertex (R) at ( 2.8,0);
        % External photons, each 0.6cm long.
        \vertex (i) at (-3.4,0);
        \vertex (f) at ( 3.4,0);

        % Open diamond endpoints on either side of the central line.
        \vertex (TL) at (-0.3, 2.5);
        \vertex (BL) at (-0.3,-2.5);
        \vertex (TR) at ( 0.3, 2.5);
        \vertex (BR) at ( 0.3,-2.5);

        % Branching point on the left and diamond attachment points on the right.
        \vertex (A) at (-1.63, 1.17);
        \vertex (J) at (-0.93, 0);
        % B makes the upper branch nearly horizontal:
        % the angle between the two left branches is about 45 degrees.
        \vertex (B) at (2.4,-0.4);
        \vertex (K) at (0.7,-2.1);

        % Upper branch: U and E lie on the straight line J--B.
        % !0.225! means 22.5% of the distance from J to B.
        \vertex (U) at ($(J)!0.225!(B)$);
        \vertex (E) at ($(J)!0.335!(B)$);

        % Lower branch: D and F lie on the straight line J--K.
        % These fractions leave a gap around the central line.
        \vertex (D) at ($(J)!0.46!(K)$);
        \vertex (F) at ($(J)!0.68!(K)$);

        % Styles: plain, photon (wavy), and gluon (spiral).
        % The asterisk keeps the vertices at the positions specified above.
        % For black internal gluons, remove the five red options.
        % The pairs J--U / E--B and J--D / F--K share common axes.
        % post length leaves 2.5mm straight segments near J.
        % The upper branch uses mirror=true, the lower branch mirror=false.
        % Diamond sides: 1.5pt; photons and gluons: 0.4pt.
        \tikzset{external line/.style={line width=1.5pt}}
        \diagram*{
            (i) -- [photon] (L);
            (R) -- [photon] (f);
            (TL) -- [plain, external line] (L) -- [plain, external line] (BL);
            (TR) -- [plain, external line] (R) -- [plain, external line] (BR);
            (J) -- [gluon, red] (A);
            (U) -- [gluon, red, decoration={mirror=true,post length=1mm}] (J);
            (D) -- [gluon, red, decoration={mirror=false,post length=1mm}] (J);
            (B) -- [gluon, red, decoration={mirror=true}] (E);
            (K) -- [gluon, red, decoration={mirror=false}] (F);
        };

        % Central line with two short horizontal segments.
        \draw [plain] (-0.38,-2.85) -- (0,-2.85)
                      -- (0,2.85) -- (0.38,2.85);
    \end{feynman}
    \end{tikzpicture}%
    }%
}
% Copy this definition into the article and call \FigIVVfour.
% Requires tikz-feynman, graphicx, and the TikZ calc library.
% The optional argument scales the entire figure: \FigIVVfour[0.7].

%----------------------------------------
% FIGURE IVV4
%----------------------------------------
\newcommand{\FigIVVfour}[1][1]{%
    \scalebox{#1}{%
    \begin{tikzpicture}[baseline={([yshift=0.3mm]c.base)}]
    \begin{feynman}
        % Same spiral dimensions as in the previous figures.
        \tikzfeynmanset{gluon/.style={
            draw=none,
            decoration={mirror},
            postaction={draw,line width=0.65pt,decorate,decoration={
                coil,aspect=0.75,
                amplitude=3.6pt,segment length=2.16mm
            }}
        }}

        % Same frame and external photons, with coordinates in cm.
        \vertex (c) at (0,0);
        \vertex (L) at (-2.8,0);
        \vertex (R) at ( 2.8,0);
        \vertex (i) at (-3.4,0);
        \vertex (f) at ( 3.4,0);
        \vertex (TL) at (-0.3, 2.5);
        \vertex (BL) at (-0.3,-2.5);
        \vertex (TR) at ( 0.3, 2.5);
        \vertex (BR) at ( 0.3,-2.5);

        % Both gluons start at A on the upper left side.
        \vertex (A) at (-1.55, 1.25);
        % Two attachment points give an opening angle of about 34 degrees.
        \vertex (C) at (-2.20,-0.60);
        \vertex (D) at (-0.65,-2.15);

        % Diamond sides: 1.5pt; photons and gluons: 0.4pt.
        % Spirals extend to the common attachment point A.
        \tikzset{external line/.style={line width=1.5pt}}
        \diagram*{
            (i) -- [photon] (L);
            (R) -- [photon] (f);
            (TL) -- [plain, external line] (L) -- [plain, external line] (BL);
            (TR) -- [plain, external line] (R) -- [plain, external line] (BR);
            (C) -- [gluon, red, decoration={mirror=false,post length=1mm}] (A);
            (D) -- [gluon, red, decoration={mirror=true,post length=1mm}] (A);
        };

        % Same central line with two short horizontal segments.
        \draw [plain] (-0.38,-2.85) -- (0,-2.85)
                      -- (0,2.85) -- (0.38,2.85);
    \end{feynman}
    \end{tikzpicture}%
    }%
}

%----------------------------------------
% FIGURE IRV8
%----------------------------------------
\newcommand{\FigIRVeight}[1][1]{%
    \scalebox{#1}{%
    \begin{tikzpicture}[baseline={([yshift=0.3mm]c.base)}]
    \begin{feynman}
        % Same spiral dimensions as in the previous figures.
        \tikzfeynmanset{gluon/.style={
            draw=none,
            decoration={mirror},
            postaction={draw,line width=0.65pt,decorate,decoration={
                coil,aspect=0.75,
                amplitude=3.6pt,segment length=2.16mm
            }}
        }}

        % Same frame and external photons, with coordinates in cm.
        \vertex (c) at (0,0);
        \vertex (L) at (-2.8,0);
        \vertex (R) at ( 2.8,0);
        \vertex (i) at (-3.4,0);
        \vertex (f) at ( 3.4,0);
        \vertex (TL) at (-0.3, 2.5);
        \vertex (BL) at (-0.3,-2.5);
        \vertex (TR) at ( 0.3, 2.5);
        \vertex (BR) at ( 0.3,-2.5);

        % Both gluons start at A on the upper left side.
        \vertex (A) at (-1.55, 1.25);
        % One attachment on the lower left side and one on the right.
        \vertex (C) at (-1.55,-1.25);
        \vertex (B) at ( 1.88,-0.92);

        % U and E lie on A--B, leaving a gap around the central line.
        \vertex (U) at ($(A)!0.400!(B)$);
        \vertex (E) at ($(A)!0.505!(B)$);

        % Diamond sides: 1.5pt; photons and gluons: 0.4pt.
        % A minimal end length keeps the first turns from overlapping.
        \tikzset{external line/.style={line width=1.5pt}}
        \diagram*{
            (i) -- [photon] (L);
            (R) -- [photon] (f);
            (TL) -- [plain, external line] (L) -- [plain, external line] (BL);
            (TR) -- [plain, external line] (R) -- [plain, external line] (BR);
            (C) -- [gluon, red, decoration={mirror=false,post length=1mm}] (A);
            (U) -- [gluon, red, decoration={mirror=true,post length=1mm}] (A);
            (B) -- [gluon, red, decoration={mirror=true}] (E);
        };

        % Same central line with two short horizontal segments.
        \draw [plain] (-0.38,-2.85) -- (0,-2.85)
                      -- (0,2.85) -- (0.38,2.85);
    \end{feynman}
    \end{tikzpicture}%
    }%
}

%----------------------------------------
% FIGURE IRR7
%----------------------------------------
\newcommand{\FigIRRseven}[1][1]{%
    \scalebox{#1}{%
    \begin{tikzpicture}[baseline={([yshift=0.3mm]c.base)}]
    \begin{feynman}
        % Same spiral dimensions as in the previous figures.
        \tikzfeynmanset{gluon/.style={
            draw=none,
            decoration={mirror},
            postaction={draw,line width=0.65pt,decorate,decoration={
                coil,aspect=0.75,
                amplitude=3.6pt,segment length=2.16mm
            }}
        }}

        % Same frame and external photons, with coordinates in cm.
        \vertex (c) at (0,0);
        \vertex (L) at (-2.8,0);
        \vertex (R) at ( 2.8,0);
        \vertex (i) at (-3.4,0);
        \vertex (f) at ( 3.4,0);
        \vertex (TL) at (-0.3, 2.5);
        \vertex (BL) at (-0.3,-2.5);
        \vertex (TR) at ( 0.3, 2.5);
        \vertex (BR) at ( 0.3,-2.5);

        % Both gluons start at A on the upper left side.
        \vertex (A) at (-1.55, 1.25);
        % Two attachment points give an opening angle of about 34 degrees.
        \vertex (B) at (2.40,-0.40);
        \vertex (K) at (0.65,-2.15);

        % Each pair of cut segments lies on a single straight line.
        % The fractions leave a gap around the central line.
        \vertex (U) at ($(A)!0.347!(B)$);
        \vertex (E) at ($(A)!0.438!(B)$);
        \vertex (D) at ($(A)!0.623!(K)$);
        \vertex (F) at ($(A)!0.786!(K)$);

        % Diamond sides: 1.5pt; photons and gluons: 0.4pt.
        % A minimal end length keeps the first turns from overlapping.
        \tikzset{external line/.style={line width=1.5pt}}
        \diagram*{
            (i) -- [photon] (L);
            (R) -- [photon] (f);
            (TL) -- [plain, external line] (L) -- [plain, external line] (BL);
            (TR) -- [plain, external line] (R) -- [plain, external line] (BR);
            (U) -- [gluon, red, decoration={mirror=true,post length=1mm}] (A);
            (D) -- [gluon, red, decoration={mirror=false,post length=1mm}] (A);
            (B) -- [gluon, red, decoration={mirror=true}] (E);
            (K) -- [gluon, red, decoration={mirror=false}] (F);
        };

        % Same central line with two short horizontal segments.
        \draw [plain] (-0.38,-2.85) -- (0,-2.85)
                      -- (0,2.85) -- (0.38,2.85);
    \end{feynman}
    \end{tikzpicture}%
    }%
}

%%%%%%%%%%%%%%%%%%%%%%%%%%%%%%%%%%%%%
%%% SECTIONS 
%%%%%%%%%%%%%%%%%%%%%%%%%%%%%%%%%%%%%

\section{Introduction}

Motivated by an exciting opportunity to
explore the fundamental laws of Nature through precision measurements at the Large
Hadron Collider (LHC), particle theorists have made remarkable progress in advancing the frontier of perturbative calculations in recent years \cite{Heinrich:2020ybq,Pellen:2024snm}. A direct manifestation of this progress is a large number of hadron collider processes for which next-to-next-to-leading order (NNLO) Quantum Chromodynamics (QCD) predictions are available \cite{Czakon:2013goa,Behring:2020uzq,Czakon:2021mjy,Asteriadis:2022ebf,Gehrmann-DeRidder:2023gdl,Badger:2023mgf,Asteriadis:2024nbg,Kim:2024kaq,Badger:2025ilt,NNLOJET:2025rno}, a growing number of processes for which next-to-next-to-next-to-leading order (N3LO) predictions have been obtained \cite{Anastasiou:2015vya,Duhr:2021vwj,Chen:2022cgv,Baglio:2022wzu}, and the appearance of many new theoretical results suggesting that an N3LO QCD description of generic processes at the LHC may actually become feasible in the not-too-distant future \cite{Caola:2021izf,Caola:2022dfa,Gehrmann:2023jyv,Baranowski:2024vxg,Chen:2025utl,Chen:2025kez}. Since these achievements are further supported  by continuous progress in the development of simulation tools and in the understanding of resummations \cite{Nason:2004rx,Monni:2019whf,Bierlich:2022pfr,Gavardi:2023aco,vanBeekveld:2024wws,Sherpa:2024mfk,Alioli:2025xcu,Generet:2025vth,Abreu:2026oxj}, there is no doubt that a highly accurate, robust and reliable description of hadron collisions will be available for the analyses of LHC data.

However, because these advances primarily concern \emph{partonic cross sections}, the extent to which they improve predictions for collisions of \emph{actual protons} depends on how accurately non-perturbative effects can be accounted for. For suitably defined observables characterized by a hard scale $Q$, non-perturbative effects are expected to induce relative corrections of order $(\Lambda_{\rm QCD}/Q)^n$, where $n$ is a positive power and $\Lambda_{\rm QCD}$ denotes the characteristic non-perturbative scale of QCD. Since in a typical collider process 
$\Lambda_{\rm QCD} \ll Q$, 
only the $n=1$ case (linear power corrections) should be considered at the  precision attained by current measurements \cite{ATLAS:2024erm,ATLAS:2024dxp,Camarda:2022qdg,Nason:2023asn} and measurements planned in the foreseeable future \cite{ATLAS:2025eii}.

In the absence of first-principles computations of non-perturbative corrections to collider processes, current studies of these effects rely on models. Some of these models -- e.g. the renormalon model \cite{Gross:1974jv,Lautrup:1977hs,tHooft:1977xjm,Parisi:1978bj,Mueller:1984vh, Beneke:1998ui} -- can be connected to certain simple quantum field theories. In this case, the problem of estimating non-perturbative effects can be reformulated as the problem of studying the infrared (IR) sensitivity to a particular parameter of the theory, typically an auxiliary gluon mass. While these studies lead to interesting results \cite{Caola:2021kzt,Caola:2022vea,Makarov:2023ttq}, they have been  limited to an Abelianized version of QCD, such that their extension to full non-Abelian QCD remains conjectural. Nevertheless, all those studies reveal a common pattern: linear power corrections arising from phase space and matrix elements \textit{cancel},
leaving observables as the only source of such 
corrections.
Establishing whether this cancellation persists beyond the Abelian approximation and also in higher perturbative orders 
is an important step toward understanding linear power corrections in QCD.

With this aim, we introduced in Ref.~\cite{Fontes:2026ihz} a framework for probing the IR sensitivity of collider processes, based on a spontaneously broken non-Abelian $\SU{2}$ gauge theory. By means of the Higgs mechanism, the gauge bosons acquire a mass $\mg$, which we take
to be small. The dependence of physical quantities on $m_g$ then serves as a measure of their IR sensitivity;
in this sense, 
we would like to 
interpret a linear dependence on $\mg$ 
in the toy model as a 
sign of linear non-perturbative power corrections to analogous observables in QCD.

Since in the toy model one can compute the dependence on $m_g$ to, in principle, arbitrary perturbative orders, it becomes possible to use   this framework to study the degree of  infra-red sensitivity of concrete processes.  With heavy-quark production in mind, as a first step, we computed through two loops the relation between the pole and $\overline{\rm MS}$ masses of a heavy quark, as well as the heavy-quark 
wave-function renormalization constant, and found a linear dependence on $m_g$ in both  quantities. The presence of such terms, expected from the known linear IR sensitivity of the pole mass
\cite{Bigi:1994em,Beneke:1994sw,Beneke:1994rs}, provided a consistency  check of the framework.

In this paper, we use the same framework to  study the cross section of heavy-quark pair production by a color-neutral current at NNLO, which is the first perturbative order sensitive to non-Abelian effects. As in earlier 
studies \cite{Makarov:2023ttq,Makarov:2023uet}, we 
 observe that  individual contributions arising from phase space and matrix elements contain terms linear in $\mg$, but these terms fully cancel in the final result provided 
 that  one uses 
 a heavy-quark mass  
 that is free of linear $m_g$ effects. 

An interesting aspect of our calculation is that we establish the cancellation in the differential cross section without explicitly evaluating any integral. Instead,  we use the residue theorem to 
relate loop and  real-emission phase-space integrals, enabling an 
algebraic cancellation  of all terms linear in $\mg$.
This use of residue relations is reminiscent of loop–tree duality \cite{Catani:2008LTD,Bierenbaum:2010LTD}, which has developed considerably in recent years, notably through its application to local unitarity \cite{Capatti:2020xjc,Capatti:2022tit}.

Although the cancellation occurs  at the level of the differential cross section, it gets  spoiled by  observables that  themselves  develop a linear sensitivity \cite{Caola:2022vea}. Identifying infrared-safe observables that preserve the cancellation is an interesting question, which  is beyond the scope of this paper.

Finally, the cancellation found here provides a concrete non-Abelian example of the proposed extension \cite{Akhoury:1995cj,Akhoury:1996ks} of the Bloch–Nordsieck \cite{Bloch:1937pw} and Kinoshita–Lee–Nauenberg \cite{Kinoshita:1962ur,Lee:1964is} theorems  
%beyond logarithmic infrared sensitivity 
to linear infrared sensitivity. Earlier calculations,  furnishing  
support for this proposal, 
dealt primarily with Abelianized  cases
\cite{Beneke:1994bc,Beneke:1995pq,Makarov:2023ttq,Makarov:2023uet} with the notable exception 
of Ref.~\cite{Sinkovics:1998mi}.
The results reported in this paper 
establish the cancellation
of linear infrared-sensitive terms 
in a particular non-Abelian theory, albeit for a 
specific collider process.

The remainder of the paper is organized as follows. After summarizing the $\SU{2}$ gauge theory proposed in Ref.~\cite{Fontes:2026ihz} in Section~\ref{sec:model}, we present an overview of  the calculation and the adopted workflow in Section~\ref{sec:3}. Section~\ref{sec:4}  introduces the calculation of the differential cross section for heavy-quark pair production in the model described in Section~\ref{sec:model}, by considering it 
at both leading order (LO) and next-to-leading order (NLO).  The latter is sufficiently simple to present in detail; we use this presentation to  illustrate the main features of our approach. We describe the general setup of the NNLO calculation
in Section~\ref{sec:5} and discuss various details  in the following sections. 
In Section~\ref{sect:MI} we explain how 
the double-virtual loop integrals can be 
related to the real-virtual and the double-real ones. In  Section~\ref{sect:7} we describe how we deal with the renormalization and the need to 
express the cross section through a short-distance mass. In 
Section~\ref{sect:8} we  
discuss in detail how the dependence of the cross section on the first power of the gluon mass $m_g$ drops out.  We conclude  in 
Section~\ref{sect:9}. Additional technical details can be found in the appendices.

\section{The model}
\label{sec:model}
As a proxy for QCD,  
we employ the renormalizable toy model  introduced and discussed in detail in Ref.~\cite{Fontes:2026ihz}.
It is a non-Abelian gauge theory based on the $\SU{2}$ gauge group, which is spontaneously broken by the Higgs mechanism.
Following the QCD-inspired terminology adopted in Ref.~\cite{Fontes:2026ihz}, we refer to the gauge fields as gluons and to the massive Dirac fermion as the top quark. As the name suggests, it plays the role of a heavy ($m_q \gg \Lambda_{\rm QCD}$) quark in the 
Standard Model.
The theory also contains $\nf$ flavors of massless quarks.%
\footnote{The existence of massless quarks implies that the (massive) gluon becomes unstable. We will see manifestations 
of this instability in the NNLO computation.}

In this section, we summarize the ingredients of the model that are needed for the calculation presented below, 
and we refer to Ref.~\cite{Fontes:2026ihz} for a more detailed discussion of its particle content, Feynman rules and renormalization.
The Lagrangian of the theory reads
\begin{equation}
\begin{split}
    \calL
    = & - \frac{1}{4} G^{a}_{\mu\nu} G^{a,\mu \nu}
    + \barpsi_t (i\slashed{D} - \mt) \psi_t 
    + \sum \limits_{k=1}^{\nf} 
    \barpsi_k i \slashed{D} \psi_k \\
    & + (D_\mu \Phi)^\dagger (D^\mu \Phi) 
    + \mu^2 \Phi^\dagger \Phi 
    - \lambda (\Phi^\dagger \Phi)^2
    + \calL_\GF + \calL_\ghost \,.
    \label{eq_Lagrangian}
\end{split}
\end{equation}
Here, $\Phi$ denotes the Higgs doublet, $\psi_t$ is the top-quark field with mass $\mt$, and $\psi_{k=1,\ldots,\nf}$ are the $\nf$ massless-quark fields.
We note that the top-quark mass is  not generated by spontaneous symmetry breaking and, therefore, is a completely independent
 parameter.
The terms $\calL_\GF$ and $\calL_\ghost$ describe the gauge-fixing and ghost sectors, respectively; their explicit expressions can be found in Eq.~(2.5) of Ref.~\cite{Fontes:2026ihz}.

The covariant derivative and the field-strength tensor are defined as
\begin{equation}
    D_\mu = \partial_\mu + i \gs \hatT^a G^a_{\mu}, 
    \qquad
    G_{\mu\nu}^a = \partial_\mu G_\nu^a - \partial_\nu G_\mu^a - \gs f^{abc} G_\mu^b G_\nu^c \,,
    \label{eq2.2}
\end{equation}
where $G_\mu^a$ is the gluon field and $\gs$ is the gauge coupling.
For $\SU{2}$, the generators and structure constants are $\hatT^a = \sigma^a/2$ and $f^{abc}=\ep^{abc}$, respectively.
Although the gauge group is fixed to $\SU{2}$, in principle we can retain the generic Casimir operators $\Ca$, $\Cf$ and $\TR$ in our results throughout the calculation, in order to keep track of the origin of the different terms in amplitudes 
and cross sections.

Spontaneous symmetry breaking is induced by the non-vanishing vacuum expectation value $v$ of the Higgs field.
We parametrize the Higgs doublet as
\begin{equation}
    \Phi(x) = \frac{1}{\sqrt{2}}
    \begin{pmatrix}
    - \varphi_2(x) - i\varphi_1(x) \\ v + H(x) + i\varphi_3(x)
    \end{pmatrix},
    \label{eq_Phi_doublet_def}
\end{equation}
where $H$ is the physical Higgs field and $\varphi_{1,2,3}$ are the would-be Goldstone bosons.
The gluons and the Higgs boson acquire masses $\mg$ and $\mh$, respectively, with
\begin{equation}
    \label{eq:parameter-relations}
    \mg = \frac{gv}{2} \,, 
    \qquad
    \mh = \sqrt{2 \lambda} \, v \,,
    \quad {\rm with} \quad
    v = \sqrt{\frac{\mu^2}{\lambda}} \,.
\end{equation}

After spontaneous symmetry breaking, the physical spectrum consists of the top quark with  mass $\mt$, $\nf$ massless quarks, three massive gluons with a common mass $\mg$, and one Higgs boson with  mass $\mh$.
Since we work in the Feynman gauge, the would-be Goldstone bosons, ghosts and antighosts with the mass $m_g$  additionally appear as unphysical degrees of freedom.  These particles 
must be included together with the gluons when considering the relevant real and virtual contributions.

Throughout the paper, we use dimensional regularization with $d=4-2\ep$ and renormalization scale $\mu$.  We introduce the renormalization of the bare coupling constant $g_{\rms}^{(0)}$ in the standard way
\be \label{eq:DefCouplingAndSDMass}
    g_{\rms}^{(0)} = g_{\rms}(\mu) S_\ep
    \mu^{\ep} \left( 1 + \frac{\as}{2 \pi}  Z_g^{(1)} + \order{\as^2} \right) ,
\ee
where $S_\ep = (4\pi)^{-\ep/2} e^{\ep \gamma_E/2}$.  We define $\as = \as(\mu) = \gs^2(\mu)/(4 \pi)$. 
Other renormalization constants of the model were computed in Ref.~\cite{Fontes:2026ihz} and we will not repeat this discussion here.
\section{Setup of the calculation}
\label{sec:3}

In this paper, we consider the \emph{exclusive} production of a heavy-quark pair $t \bar{t}$ by a neutral vector current through ${\cal O}(\as^2)$ in the model described above.
The cross section is a function of the energy of the process $\sqrt{s}$, the top-quark mass $\mt$, the gluon mass $m_g$, and a resolution parameter $Q_{\rm veto} \gg m_g$ which defines the exclusivity of the process. One can ask about $Q_{\rm veto}$-dependent linear terms at ${\cal O}(\as^2)$ -- a question that is best explored in the context of power corrections 
to $t \bar{t} + j$ production, which we do not study  in this article. As a consequence, we will ignore the situation in which any of the gluons in the final state has an energy that is comparable to $Q_{\rm veto}$, thus significantly exceeding $m_g$.
Our goal is to show that the 
$Q_{\rm veto}$-independent part of the exclusive  cross section, through second order in the 
expansion in $\as$, does not contain terms linear in $m_g$.

To this end, we write the perturbative expansion of the differential cross section as 
\be
\label{eq:superscripts}
{\rm d} \sigma_{t \bar t}
= {\rm d}\sigma^{(0)}_{t \bar t} 
+ \left ( \frac{\as}{2 \pi}
\right ) {\rm d} 
\sigma^{(1)}_{t \bar t} 
+ \left ( \frac{\as}{2 \pi}
\right )^2 {\rm d} 
\sigma^{(2)}_{t \bar t}+ \mathcal{O}(\as^3) .
\ee
We use this notation for an expansion in $\as/(2 \pi)$ throughout the paper also for other quantities.
To compute ${\rm d} 
\sigma^{(1,2)}_{t \bar t} $, 
we use the well-developed machinery of multi-loop computations in two independent implementations.%
\footnote{We resort to the publicly available codes FeynMaster \cite{Fontes:2019wqh,Fontes:2021iue,Fontes:2025svw},
Qgraf~\cite{Nogueira:1991ex}, FeynCalc~\cite{Mertig:1990an,Shtabovenko:2016sxi,Shtabovenko:2020gxv,Shtabovenko:2023idz,Shtabovenko:2025lxq}, FeynHelpers \cite{Shtabovenko:2016whf,Shtabovenko:2025lxq}
and FIRE~\cite{Smirnov:2025prc}.}
Starting from full diagrams, we apply a transformation that maps real-emission phase spaces onto the Born phase space, which we discuss further below. Subsequently we construct the expansion of real and virtual graphs in the gluon mass using the strategy of regions \cite{Beneke:1997zp}. We then use the integration-by-parts (IBP) technology \cite{Chetyrkin:1981qh}, together with reverse unitarity \cite{Anastasiou:2002yz}, 
to express virtual and real integrals through a set of master integrals. Combining them, we obtain the ${\cal O}(m_g)$ correction to the cross section expressed through master integrals. 

These master integrals are, however, rather intricate to compute, especially those that simultaneously depend on the momenta of 
$t$ and $\bar{t}$. Hence, we found it convenient to rewrite loop  master integrals through integrals that describe the real emission of soft gluons.  
As mentioned in the Introduction, this rewriting resembles loop–tree duality, which expresses virtual integrals in terms of on-shell cut contributions \cite{Catani:2008LTD,Bierenbaum:2010LTD}. 
These aspects will be discussed in more detail in Section~\ref{sect:MI}. 
%-----------------------------------------------------
% NLO
%-----------------------------------------------------
\section{Calculations at leading and next-to-leading order}
\label{sec:4}

\begin{figure}[t]
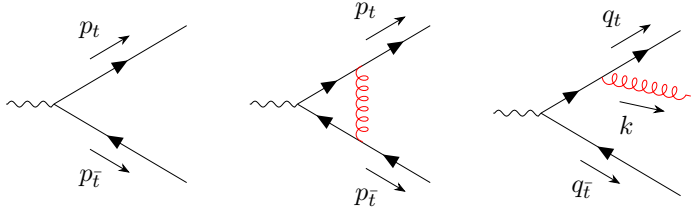

  \centering
  \TopPairTree[0.90]\hspace{1.8em}
  \TopPairVirtual[0.90]\hspace{1.8em}
  \TopPairReal[0.90]
  \caption{Tree-level, virtual and real-emission contributions.}
  \label{fig:top-pair-three}
\end{figure}

The LO matrix element for $t \bar{t}$ production by a neutral vector current  reads
\be
{\cal M}^{(0)} = \bar u(\pt) \gamma^\mu v (\ptb) 
\, \epsilon_\mu,
\ee
with $\pt$ and $p_{\bar t}$ being the momenta of the top quark and anti-top quark, respectively,  and $\epsilon^\mu$ describing 
a generic ``polarization state'' of the current. 
The corresponding diagram is shown 
in the left panel of Figure~\ref{fig:top-pair-three}.

We define $P \equiv \pt + \ptb$, such that $P^2 = s$. Since  the vector current is conserved, we will use  $\epsilon_\mu \epsilon^*_\nu \to -g_{\mu \nu}$ in what follows.\footnote{
We note that to obtain the actual cross section 
for $e^+e^- \to t \bar t$ one will have to express $\epsilon_\mu \epsilon_\nu^*$ through two independent tensor structures that describe the leptonic tensor,  with $g_{\mu \nu}$ being just one of them.   For simplicity, we will ignore the second one in what follows. However, we emphasize that we have checked that our conclusion about 
the absence of linear ${\cal O}(m_g)$ 
effects applies to both tensor structures. 
} 
Computing the differential cross section,
we find
\be \label{eq:tree-level}
\frac{ {\rm d} \sigma_{t\bart}^{(0)} }{{\rm d} \Omega_t}
=  \sigma_0
\beta^{1-2\ep} \left ( 3 - 2 \ep - \beta^2 \right ).
\ee
Here, $\sigma_0$ is  a normalization factor, $\beta = \sqrt{1-4 \mt^2/s}$ is the velocity of the heavy  quark and the solid angle ${\rm d} \Omega_t$ describes the direction of the outgoing quark in the rest frame of the vector $P^\mu$. We note that 
for the $t \bar t$ final state, the following identities hold
\be
    \pt \cdot \ptb = \frac{s}{4}(1 + \beta^2) \,,
    \qquad
    \mt^2 = \frac{s}{4}(1 - \beta^2) \, .
\ee

We are interested in computing radiative corrections 
to the cross section in Eq.~(\ref{eq:tree-level}) and studying the limit $m_g \to 0$ focusing 
on ${\cal O}(m_g)$ terms.
At NLO, we have to combine the standard ingredients, i.e., one-loop virtual corrections, the top-quark wave-function renormalization constant $Z_{\psit}$, and real-emission corrections. 
Furthermore, as will be discussed in detail, we also have to express the NLO contribution to the cross section in terms of the short-distance heavy-quark mass, since only in this case do the NLO linear mass corrections cancel \cite{Makarov:2023uet}.

In principle, one can compute all the required ingredients 
exactly and then expand in $m_g \ll \sqrt{s}, \mt$.
This, however, will become impossible beyond the NLO. 
Hence, we would like to organize the calculation in such a way that the expansion in $m_g$ 
is performed  \emph{before} the integration over momenta of both 
virtual and real gluons. A procedure 
to do that was discussed in Refs \cite{Makarov:2023uet,Fontes:2026ihz}
and we  adapt  it to the present case. 

We begin with the virtual corrections. If  we adopt the on-shell scheme for the renormalization of the top-quark mass, the only diagram that needs to be considered is the vertex correction, shown in the central panel of Figure~\ref{fig:top-pair-three}.
The linear power correction 
in this case originates from the region where the loop momentum is comparable to $m_g$; we 
refer to such loop momenta as \textit{soft}.%
\footnote{We will refer to a complementary momentum 
region $k \sim m_t$ 
as \textit{hard}. This region cannot generate linear power corrections.}
Expanding the vertex correction to linear order 
in the soft momentum, and applying IBP identities, we find that the ${\cal O}(m_g)$ contribution to the one-loop vertex is expressed through the following master integral\footnote{A similar integral where $p_t$ is replaced with $- p_{\bar t}$
is equal to $I_{\rm V}$ since  $t$ and $\bar t$ have 
equal energies and are back-to-back.
}
\begin{align}    
    I_{\rm V} \equiv \int \frac{\dd[d]{k}}{i (2\pi)^d} \, \frac{1}{(k^2 - \mg^2 + \iZero)} \frac{1}{(2 k \cdot p_t + \iZero)} \, ,
\end{align}
with $\pt^2 = \mt^2$.
As suggested above, we proceed by rewriting $I_\rmV$ in terms of real-emission integrals. We start by integrating over $k^0$ using the residue theorem. Closing the integration contour in the upper half-plane (so that the pole of the quark propagator is avoided), we find 
\begin{equation}
    I_\rmV
    = \int \frac{\dd[d]{k}}{(2\pi)^{d-1}} \, \delta_+(k^2-\mg^2) \, \frac{1}
    {2 k \cdot \pt} \,,
    \label{eq5.2}
\end{equation}
where $\delta_+(k^2-\mg^2) = \theta(k^0)\delta_+(k^2-\mg^2)$.
We note that this integral is a 
\emph{real-emission} integral, since the 
positive-energy gluon is on the mass shell. Furthermore, as we will see shortly, it is the only master integral that appears when 
the real-emission contribution is considered. Since the NLO renormalization constants also  
depend on $I_\rmV$, we conclude  that \emph{all} the different contributions to the NLO differential cross section can be written in terms of the real-emission master integral, which we refer to as $I_\rmR$, 
\begin{equation}
    I_\rmR \equiv \int \frac{\dd[d]{k}}{(2\pi)^{d-1}} \, \delta_+(k^2-\mg^2) \, \frac{1}
    {2 k \cdot \pt} \,.
\end{equation}
We conclude that the linear terms in $\mg$ coming from the virtual corrections read
\begin{align}
    \mathcal{T}_{\mg} \Bigg[\frac{ {\rm d} \sigma_{t\bart,\rmV}^{(1)} }{{\rm d} \Omega_t}\Bigg] = {\cal N} \sigma_0 \Cf \beta^{-1-2\ep} (3-2\ep) 
    \big[ 1 - 2 \ep + 2\beta^2(2-\ep) - \beta^4 \big] I_\rmR  \,,
    \label{eq4.8}
\end{align}
where ${\cal T}_{m_g}$ is  an operator that extracts the ${\cal O}(m_g)$ terms from an expression that it acts upon and ${\cal N} = 
(4 \pi)^2 S_\ep^2 \mu^{2\ep}
$ is a normalization constant. 
We note that in Eq.~(\ref{eq4.8}) the $m_g$-dependence 
of the right-hand side is 
implicit because $I_\rmR \sim m_g^{1-2\ep}$.
\\

We now turn to the real-emission  
contributions, which involve the  
final state $t \bar t g$, see the right panel of Figure~\ref{fig:top-pair-three}.  In this case, we denote the momenta of the top  quark and anti-top quark as $q_t$ and $q_{\bar t}$, respectively, 
and the momentum of the gluon as 
$k$. 
The phase space of the $t \bar t  g$ final state is different 
from that of the Born process.   Yet, as 
pointed out in Ref.~\cite{Makarov:2023uet}, 
assuming that the gluon momentum 
is soft ($k \sim m_g$), 
one can map the real-emission phase space to the Born phase space with ${\cal O}(m_g)$ accuracy.
This is readily done by redefining the top-quark momenta \cite{Makarov:2023uet}
\begin{equation} \label{eq:RealToBornDefinition}
\begin{aligned}
    \qt &= \pt - \frac{1+\rbchi}{2} \, k + A\,\pt + B\,\ptb \, , \\[.2em]
    \qtb &= \ptb - \frac{1-\rbchi}{2} \, k - A\,\pt - B\, \ptb \, ,
\end{aligned}    
\end{equation}
where $A$ and $B$ are functions of 
$p_t, p_{\bar t}$ and $k$ that scale 
as the first power of $k$ and $\chi$ is an arbitrary parameter. 
Requiring
\be
     \qquad \qt^2 = \pt^2 \, , 
     \qquad  \qtb^2 = \ptb^2 \, ,
     \ee
we find 
\begin{align}
A &=
-\frac{
(1+\rbchi)\,\ptb^2\,(\pt \cdot k)
+(1-\rbchi)\,(\pt \cdot \ptb)\,(\ptb \cdot k)
}{
2\left[(\pt \cdot \ptb)^2-\pt^2\ptb^2\right]
} + {\cal O}\!\left(k^2\right) \, ,
\label{eq:RealToBornA}
\\[2mm] 
B &= -A(\rbchi \to -\rbchi,
         \pt \leftrightarrow \ptb)\, .
\end{align}
Under the mapping shown in Eq.~\eqref{eq:RealToBornDefinition}, the phase space transforms as\footnote{\eq\eqref{eq:RealToBornJacobian} agrees with Eq.~(4.7) in Ref.~\cite{Makarov:2023uet} in $d=4$ dimensions apart from a factor of $2\pi$ that is missing there.}
\begin{align}
    &\; {\rm dLips}(P; \qt,\qtb,k)
    =
    {\rm dLips_{LO}}(P; \pt,\ptb)\,
    \frac{{\rm d}^d k}{(2\pi)^{d-1}} \,\delta_+(k^2-\mg^2)
    \nonumber \\
    & \times \bigg \{
    1 +\frac{1}{(\pt \cdot \ptb)^2-\mt^4}
    \bigg[
    -(1-\epsilon)\mt^2
    \Big[
    (1+\rbchi)(\pt \cdot k)
    +(1-\rbchi)(\ptb \cdot k)
    \Big]
    \label{eq:RealToBornJacobian} \\
    &
    +(\pt \cdot \ptb)
    \Big[
    \left[\epsilon-\rbchi(1-\epsilon)\right](\pt \cdot k)
    +\left[\epsilon+\rbchi(1-\epsilon)\right](\ptb \cdot k)
    \Big]
    \bigg]+\order{k^2}
    \bigg\}
    \, .
    \nonumber 
\end{align}
We note that, since the parameter $\rbchi$ remains arbitrary, it  should drop out of the  final results.
Furthermore, ${\cal O}(k^2)$ terms are not relevant for the NLO calculation but turn out to be helpful  at NNLO. 
A more general discussion of the momentum transformation that allows one to extend the above  formulas to include 
$\order{k^2}$ and higher terms can be found in Appendix \ref{appendix:kSqRealToBorn}. 

Since the mapping in 
Eq.~\eqref{eq:RealToBornDefinition}  factorizes the $t \bar t g$ phase space into the Born phase space and the phase space of a soft on-shell gluon, one can integrate over soft-gluon momentum  
while keeping $p_t$ and $p_{\bar t}$ fixed. In practice, we use 
reverse unitarity and IBP identities to express all NLO real-emission 
integrals through a single master integral, $I_\rmR$. The result for the linear terms in $\mg$ coming from the real-emission contribution reads 
\begin{align}
    \mathcal{T}_{\mg}  \Bigg[\frac{ {\rm d} \sigma_{t\bart,\rmR}^{(1)} }{{\rm d} \Omega_t}\Bigg] = {\cal N} \sigma_0  \Cf \beta^{1-2\ep} (3-2\ep)\left ( -3 + 2 \ep + \beta^2 \right ) I_\rmR  \, .
\end{align}

As the last of the standard ingredients, we have to include the on-shell wave-function renormalization constant of the heavy quark, $Z_{\psit}$.
We find
\begin{align}
    \mathcal{T}_{\mg} \Bigg[\frac{ {\rm d} \sigma_{t\bart, \rm Ren}^{(1)} }{{\rm d} \Omega_t}\Bigg] = 
     \mathcal{T}_{\mg} \left [2 Z_{\psit}^{(1)} \frac{ {\rm d} \sigma_{t\bart}^{(0)} }{{\rm d} \Omega_t}%{\rm d} \sigma_{t\bart}^{(0)}
    \right ] = 
   {\cal N} \sigma_0 \, \Cf \beta^{1-2\ep} (3-2\ep)\left ( -3 + 2 \ep + \beta^2 \right )I_\rmR  \, ,
\end{align}
where we have used \cite{Fontes:2026ihz}
\be
\mathcal{T}_{\mg}\Big[
Z_{\psi_t}
\Big ] = 
\mathcal{T}_{\mg}  \Big[1 + \frac{\alpha_s}{2 \pi} \; Z^{(1)}_{\psit}
+{\cal O}(\alpha_s^2) 
\Big] = - \frac{{\cal N}}{2} \Cf (3-2\ep)\; \frac{\alpha_s}{2 \pi} I_\rmR +{\cal O}(\alpha_s^2) \,.
\ee

The sum of the three contributions computed above evaluates to 
\be
\mathcal{T}_{\mg} \Bigg[\frac{ {\rm d} \sigma_{t\bart,\rmV}^{(1)} }{{\rm d} \Omega_t} + \frac{ {\rm d} \sigma_{t\bart,\rmR}^{(1)} }{{\rm d} \Omega_t} + \frac{ {\rm d} \sigma_{t\bart,\rm Ren}^{(1)} }{{\rm d} \Omega_t}\Bigg] = {\cal N} \sigma_0 \Cf \beta^{-1-2\ep} (3-2\ep)(1-\beta^2)\left ( 1 - 2 \ep - \beta^2 \right )I_\rmR \,,
\label{eq:NLO:intermediateResult}
\ee
and is different from zero.
The reason is that the above result corresponds to the on-shell renormalization of the top-quark mass. 
To appreciate it, we note that
Eq.~(\ref{eq:NLO:intermediateResult})  can be written as the full derivative 
of the leading-order cross section 
with respect to $m_t^2$
\be
\mathcal{T}_{\mg} \Bigg[\frac{ {\rm d} \sigma_{t\bart,\rmV}^{(1)} }{{\rm d} \Omega_t} + \frac{ {\rm d} \sigma_{t\bart,\rmR}^{(1)} }{{\rm d} \Omega_t} + \frac{ {\rm d} \sigma_{t\bart,\rm Ren}^{(1)} }{{\rm d} \Omega_t}\Bigg] =-2{\cal N} 
 \Cf m_t^2 I_\rmR \;
\frac{{\rm d} }{{\rm d} m_t^2 }
\left [
\frac{  {\rm d} \sigma_{t\bart}^{(0)} }{ {\rm d} \Omega_t } 
\right ]
%\; 4 \Cf I_\rmR
\, .
\label{eq4.17}
\ee
Hence, we conclude that the  
${\cal O}(\mg)$ term that survives   in the sum of 
the virtual, real and the renormalization contributions 
can be removed by computing the LO cross section with a mass parameter that differs from the 
pole mass.  

We will now show that the mass shift that is  needed to remove the ${\cal O}(m_g)$ 
dependence in Eq.~(\ref{eq4.17})  is \emph{universal}, i.e.\ it can be computed independently  of a particular process. 
Indeed, it is well known that the quark pole mass develops a linear infrared sensitivity \cite{Bigi:1994em,Beneke:1994sw,Beneke:1994rs,Fontes:2026ihz}. For our model, this implies the  
appearance of 
an ${\cal O}(m_g)$ term in the relation between  the pole and any short-distance (e.g.\  $\overline{\rm MS}$) 
mass of the top quark \cite{Fontes:2026ihz}.
Therefore, we write 
\be
\label{eq:mass-rel}
\mt = \mtt + \delta \mt \, ,
\ee
where $\mtt$ is the short-distance mass 
and  $\delta m_t$ contains $ {\cal O}(\alpha_s m_g)$ terms. We
note that these terms are universal in the sense that  
%$\delta m_t$ 
they are 
independent of 
$\mtt$. 
The Born cross section in Eq.~\eqref{eq:tree-level} depends implicitly on $\mt$ via the velocity $\beta$. Inserting the mass shift given in Eq.~\eqref{eq:mass-rel}
into that equation, we 
expand 
in $\delta m_t$ to first order and obtain
\be
    \mathcal{T}_{\mg} \Bigg[\frac{ {\rm d} \sigma_{t\bart,\rm mass}^{(1)} }{{\rm d} \Omega_t}\Bigg] = - {\cal N} \sigma_0 \, \Cf \tilde{\beta}^{-1-2\ep} (3-2\ep) \big(1-\tilde{\beta}^2 \big) \big( 1 - 2 \ep - \tilde{\beta}^2 \big)I_\rmR \, ,
\label{eq4.19}
\ee
where we have used \cite{Fontes:2026ihz}
\be
\mathcal{T}_{\mg}\Big[
\delta \mt \Big] =
%\frac{2 \alpha_s   \Cf}{\pi}
\frac{\alpha_s}{2 \pi} {\cal N}
\Cf \mtt I_\rmR \, 
+{\cal O}(\alpha_s^2).
\ee
Furthermore, in Eq.~(\ref{eq4.19}) we  have also introduced the modified velocity  $\tilde{\beta}$ which, in contrast to  $\beta$,  depends on  $\mtt$ instead of $\mt$.

Since $\beta = \tilde{\beta} + {\cal O}(\as)$, we can identify $\beta$ with $\tilde{\beta}$ in \eq\eqref{eq:NLO:intermediateResult}, obtaining  a cancellation of the terms linear in $\mg$ at NLO,
\be \label{eq:NLO:finalResult}
\mathcal{T}_{\mg} \Bigg[\frac{ {\rm d} \sigma_{t\bart,\rmV}^{(1)} }{{\rm d} \Omega_t} + \frac{ {\rm d} \sigma_{t\bart,\rmR}^{(1)} }{{\rm d} \Omega_t} + \frac{ {\rm d} \sigma_{t\bart,\rm Ren}^{(1)} }{{\rm d} \Omega_t} + \frac{ {\rm d} \sigma_{t\bart,\rm mass}^{(1)} }{{\rm d} \Omega_t}\Bigg] = 0 \, .
\ee
We stress that this cancellation was obtained without the need to compute \textit{any} integral. In fact, we only had to  relate the  virtual master integral ($I_{\rm V}$) to the real-emission master integral ($I_{\rm R}$) by means of the residue theorem. The cancellation in Eq.~\eqref{eq:NLO:finalResult} occurs because the coefficients of $I_{\rm R}$ cancel each other.

%%%%%%%%%%%%%%%%%%%%%%%%%%%%%%%%%%%%%%%%%%%%%%%%
% NNLO
%%%%%%%%%%%%%%%%%%%%%%%%%%%%%%%%%%%%%%%%%%%%%%%%
\section{The general strategy of the NNLO calculation}
\label{sec:5}

The cancellation mechanism described in the previous section 
is well known in the context of 
QCD  renormalon calculus (see e.g. Refs  \cite{Beneke:1994bc,FerrarioRavasio:2018ubr,Makarov:2023ttq, Makarov:2023uet}). However, such NLO computations are essentially Abelian, 
as they are not sensitive  to the  non-Abelian interactions that are critical for QCD.  Hence, it is important to understand if   the same cancellation mechanism remains at work when non-Abelian interactions are present and when higher perturbative orders are considered. 
This is the purpose of the NNLO computation 
that we describe in this section. 

To  compute   the $\mathcal{O}(\mg)$ terms in the NNLO corrections 
to the production of a $t \bar t$  pair by a neutral vector current, 
one needs to calculate the two-loop virtual corrections to the
$t \bar t$ final state, 
account for the one-loop corrections to the $t \bar t +g$
final state, include  the final states with two additional partons, e.g.\ $t \bar t gg$, perform the renormalization at the
two-loop level, and express the Born and one-loop cross sections through short-distance masses. In all these contributions, we are only interested in the $\mathcal{O}(\mg)$ terms, since our goal is to investigate whether these terms cancel when such contributions are combined.
At the same time, it is clear that performing the required calculations first and studying the limit $m_g \to 0$ 
afterwards 
is not possible,  as many massive particles make the calculation 
technically unfeasible.   Thus, we adopt the same strategy as in the NLO case, which amounts to constructing the expansion of the virtual and real integrals in powers of $m_g$ prior to  the integration over momenta of real or virtual partons.

The double-virtual contribution can be calculated in analogy to the virtual contribution at NLO using the approach based on  the strategy of regions \cite{Beneke:1997zp}. As discussed above, for each loop integral there 
are two momentum regions: soft, 
where $k \sim m_g$, and hard, with 
$k \sim \mt$. Hence, momentum regions in the double-virtual contribution that can, potentially, give ${\cal O}(\mg)$ contributions are soft-soft, soft-hard and hard-soft, while hard-hard regions cannot produce ${\cal O}(\mg)$ terms. 

Similarly, real-virtual corrections  can be treated in analogy to the real emission at NLO.
Here the virtual gluons  can be either hard or soft, while the real emission is always considered to be soft. Before the expansion in certain regions is performed, we  absorb the 
final-state gluon momentum  into the momenta of the final-state 
quarks  using Eq.~\eqref{eq:RealToBornDefinition}.
We note that  diagrams with the self-energy insertions into 
quark propagators that are slightly 
off-shell produce  spurious ${\cal O}(\mg^{-1})$  contributions. 
Such contributions  
cancel when the (mass) renormalization of the tree-level  real-emission  contribution is included.  Since, for technical reasons, we prefer to treat  
 loop and renormalization 
terms separately  until the end of the calculation,
%Terms of ${\cal O}(\mg^{-1})$ %lead to ill-defined terms if
we find it convenient to 
extend the momentum transformation
in the  real-emission contributions discussed
in Section~\ref{sec:4}
to higher orders 
in the soft momentum $k \sim m_g$. The corresponding discussion can be found in 
Appendix~\ref{appendix:kSqRealToBorn}.

The emission of two additional partons first appears at NNLO and constitutes the double-real contribution. 
We consider both emitted partons to be soft and use  the momentum transformation formulas in \eq\eqref{eq:RealToBornDefinition}, where we set   $k \to k_1 + k_2$, with $k_{1,2}$ being the momenta of the soft  partons. 
This is possible since we only need to  work with  $\order{k}$ accuracy, in which case $\order{k^2}$ contributions play no role.
We note that we work in Feynman gauge, which entails the need to consider  the emission of pairs of gluons,   Goldstone bosons and ghosts.
Furthermore, we also need to include  the emission of pairs of massless quarks and antiquarks.

Upon performing the IBP reduction of the double-virtual, real-virtual and double-real contributions, we express the NNLO contribution to the differential cross section  in terms of the master integrals listed in Table~\ref{tab_master_integrals}  in Appendix~\ref{appendix_sec_list_MIs}. 
Focusing on the contribution from the soft-soft region, we find that there are ten double-virtual, seventeen real-virtual and fourteen double-real integrals 
that have to be evaluated  
to obtain the cross section with ${\cal O}(m_g)$ accuracy. 

There are different ways to compute these integrals. One possibility  would involve a derivation of a system of coupled  differential equations that these integrals satisfy, followed by solving such a system either  analytically or numerically. 
However, we have seen that  establishing the cancellation 
of ${\cal O}(m_g)$ terms in the NLO case  
did not require us to compute 
\emph{any} integral; rather, it was important to express all integrals that appear in the virtual and real contributions, as well as in the renormalization terms, through the minimal number of  independent integrals, which in the NLO case turned out to be   one.  

We will follow the same idea in the NNLO case.  
Although, for obvious reasons,  implementing it turns out to be more difficult,  we have found 
no showstoppers. In the next section, we explain how the double-virtual integrals can be expressed through  the real-virtual and double-real ones by integrating over the energy components of virtual gluons.  
We consider two representative examples in Sections \ref{sec:case1} and \ref{sect:ivv1}, and summarize the results for the remaining master integrals in \Sec\ref{subsec_results_Ivv}.
%================================================
% MAPPING DOUBLE-VIRTUAL
%================================================
\section{Rewriting the soft-soft double-virtual integrals}
\label{sect:MI}

In this section, we explain how to 
connect double-virtual soft-soft integrals with the double-real and real-virtual ones. Since for the  two-loop integrals   the required manipulations become considerably more involved,
we begin by introducing notation that allows us to keep the calculations compact. 
%================================================
% NOTATION
%================================================
\subsection{Notation}
\label{sec:notation}
We consider soft-soft two-loop integrals describing the production of a quark and an antiquark with four-momenta $\pt$ and $\ptb$, respectively.
We write these momenta as
\begin{equation}
    \pt^\mu = E_t \, \Fm^\mu,
    ~ \text{with} ~ \Fm^\mu = (1,  \vec\Fm) \,,
    \qquad
    \ptb^\mu  = E_{\bart} \, \Sm^\mu,
    ~ \text{with} ~ \Sm^\mu = (1, \vec\Sm)  \,.
\label{eq_amu_bmu_def}
\end{equation} 
The three-vectors $\vec\Fm, \vec\Sm$ are the velocity vectors of the quark and antiquark.\footnote{We note that in the center of mass frame $\vecu = -\vecr$.}
Since all soft-soft virtual integrals are homogeneous under rescalings of the four-vectors $p_t^\mu$ and $p_{\bart}^\mu$, and since the quark energies are positive, the factors $E_t$ and $E_{\bart}$ can be factored out of the integrals 
and restored later.
The four-vectors $\Fm^\mu$ and $\Sm^\mu$ can therefore be used directly in the integrals.
We also define
\begin{gather}
k_i^\mu  = (\om_i, \veck_i) \,,\qquad
    \om_i = \sqrt{\veck_i^2 + m_i^2} \,,
    \qquad
    k_{12}^\pm = k_1 \pm k_2 \,,
\end{gather}
and denote the regular and eikonal (massive-fermion) propagators as 
\begin{equation}
    G_x(k) = \frac{1}{k^2 - m_x^2 + \iZero} \,,
    \qquad
    F_p(k) = \frac{1}{2 k\cdot p + \iZero} \,.
\label{eq_G_F_defs}
\end{equation}
The four-vector $p$ is one of the  external four-momenta ($u$ or $r$), while the subscript $x$ in $G_x(k)$ indicates the mass of the particle associated with that propagator.
The soft master integrals listed in Table~\ref{tab_master_integrals} contain $G_g$ for gluons and $G_0$ for massless quarks and, for simplicity, we omit the subscript $g$ from $G_g$ in what follows.
For real momenta, the propagators defined in \eq\eqref{eq_G_F_defs} satisfy the following identities,
\begin{equation}
    G_x(k) = G_x^*(k) - 2\pi  i\, \delta(k^2 - m_x^2),
    \qquad
    F_p(\pm k) = - F_p^*(\mp k) \,,
\label{eq:GF-identities}
\end{equation}
which we use in the examples below.

It is convenient to denote the soft-soft double-virtual integrals in the following way
\begin{equation}
   \left\langle \calI(k_1,k_2) \right\rangle  \equiv  \int\! \frac{\rmd^d k_1}{ i(2 \pi)^d } 
    \int\! \frac{\rmd^dk_2}{ i(2 \pi)^d} \; \calI(k_1,k_2) \,.
        \label{eq6.5}
\end{equation}
A typical integrand ${\cal I}(k_1,k_2)$ may contain gluon, massless-quark and eikonal propagators.
As we saw, our aim is to map such double-virtual integrals onto the real-virtual and double-real integrals.
We achieve this by integrating over the temporal components of the loop momenta $k_{1,2}$ in Eq.~(\ref{eq6.5}) using  Cauchy's theorem. Although the calculation of the residues  is elementary, it is useful to introduce a few relations that simplify the required algebra.

We start by addressing residues of gluon propagators. 
Consider an integrand $\calI(k_1,k_2)$ containing a gluon propagator $G(k_1)$,
\begin{equation}
    \calI(k_1,k_2) = G(k_1) \calR(k_1,k_2) \,.
\label{eq_6.6}
\end{equation}
To use  Cauchy's theorem to integrate  ${\cal I}(k_1,k_2)$ over $k_{1}^0$, we need to compute the residue of ${\cal I}(k_1,k_2)$ at the poles of $G(k_1)$, which are
\begin{equation}
    k_1^0 = -\omega_1 + \iZero \,,
    \qquad 
    k_1^0 = \omega_1 - \iZero \,.
\end{equation}
Depending on the global properties of ${\cal I}(k_1,k_2)$, it may be convenient to close the integration contour in either the upper half-plane ($\text{\small UP}$) or the lower half-plane ($\text{\small LP}$) of the complex $k_1^0$-plane.
If we close the integration contour in the upper (lower) half-plane, we have to pick up the first (second) pole. 

Focusing first on the contour closed in the upper half-plane, we find
\begin{equation}
    \text{Res}\big|_{(k_1^2 = \mg^2)}^{\text{\tiny UP}} \lint G(k_1) \calR(k_1,k_2) \rint
    =  - \left\langle \calR(-k_1,k_2) \right\rangle_1 \,.
\label{eq_Res_G_UP_rule}
\end{equation}
The subscript 1 on the right-hand side indicates that the gluon with momentum $k_1$ is on the mass shell, i.e., 
\begin{equation}
\label{eq:6.9}
    \lint \calR(k_1,k_2) \rint_1
    \equiv
    \int\! \frac{\rmd^dk_1}{(2 \pi)^{d-1}} \, \delta_+(k_1^2 - \mg^2)
    \int\! \frac{\rmd^dk_2}{ i(2 \pi)^d} \, \calR(k_1,k_2) \,.
\end{equation}
Therefore, the \rhs\ of \eq\eqref{eq_Res_G_UP_rule} is such that the remaining $k_1$-integral is a real-emission integral.
Note that, in deriving \eq\eqref{eq_Res_G_UP_rule}, we replaced $\vec k_1 \to -\vec k_1$ in order to construct the four-vector $-k_1$ from $(-\omega_1, \vec k_1)$, which emerges when the residue in the upper half-plane of the complex $k_1^0$-plane is taken. 
This parity transformation can be applied provided that the additional imaginary part in $k_1^0$ plays no role after the residue is taken. 
While this is true for the majority of the integrals that we need, some cases require additional care. 

We return to the integrand in \eq\eqref{eq_6.6} and note that  if the contour is closed in the lower half-plane, an analogous calculation gives
\begin{equation}
	\text{Res}\big|_{(k_1^2 = \mg^2)}^{\text{\tiny LP}} \lint G(k_1) \calR(k_1,k_2) \rint
	=  - \left\langle \calR(k_1,k_2) \right\rangle_1 \,.
	\label{eq_Res_G_LP_rule}
\end{equation}
As in the case of the upper half-plane, when this expression is used  in the presence of additional gluon or eikonal propagators, an ambiguity arises if the induced and existing $\iZero$ prescriptions have opposite signs. In this case,  additional care is needed.
One such case is the integral $\IVV{1}$, which is discussed in Section~\ref{sect:ivv1}.

Another set of useful relations involves the residues of the eikonal (top-quark) propagators.  To describe them, we consider an integrand $\calI(k_1,k_2)$ containing an eikonal propagator $F_p(\pm k_1)$, i.e.
\begin{equation}
    \calI(k_1,k_2)
    = F_p( \pm k_1) \calR(k_1,k_2) \,.
\end{equation}
The four-vector $p$ is one of the two vectors $\Fm$ or $\Sm$, introduced above.
The eikonal propagator has a single pole, which can lie either above or below the real axis.
It follows that
\begin{equation}
\begin{aligned}
    F_p(-k_1) \implies
    & \text{upper half-plane: }
    k_1^0 = \veck_1 \cdot \vecp + \iZero \,, \\
    F_p(k_1) \implies
    & \text{lower half-plane: } \hspace{1.2mm}
    k_1^0 = \veck_1 \cdot \vecp - \iZero \,.
\end{aligned}
\end{equation}
Repeating the steps used for the gluon propagator, we obtain
\begin{equation}
    \text{Res}\big|_{F_p(-k_1)}^{\text{\tiny UP}} \! \lint F_p(-k_1) \calR(k_1,k_2) \rint
    = \text{Res}\big|_{F_p(k_1)}^{\text{\tiny LP}} \! \lint F_p(k_1) \calR(k_1,k_2) \rint
    = - \frac{1}{2} \left\langle \calR(k_{1,p},k_2) \right\rangle_{\vec 1} \,,
\label{eq_fermion_pole_UP_LP_plane}
\end{equation}
where we introduce the notation
\begin{equation}
    k_{1,p} \equiv (\veck_1 \cdot \vecp, \veck_1) \,,
    \qquad
    \langle \calR(k_{1,p},k_2) \rangle_{\vec 1} \equiv \int \frac{\rmd^{d-1}\veck_1}{(2 \pi)^{d-1}} 
    \int \frac{\rmd^{d}k_2}{i(2 \pi)^{d}} \, 
    \calR(k_{1,p},k_2) \,.
\label{eq_k_i_p_def}
\end{equation}
As in the case of the gluon propagator, \eq\eqref{eq_fermion_pole_UP_LP_plane} may potentially be ill-defined because of the sign of the $ \iZero $ prescription.
However, for the integrals $\IVV{1-10}$ relevant to our calculation, it is always possible to avoid such a situation.

%================================================
% IVV{4}
%================================================
\subsection{An integral with gluon and eikonal poles}
\label{sec:case1}
As a first example of a double-virtual integral, we consider $\IVV{4}$ (cf.\ Table \ref{tab_master_integrals}),
\begin{equation}
    \IVV{4}
    = \vcenter{\hbox{\FigIVVfour[0.35]}}
    = \lint G(k_1) G(k_2) F_\Sm(-k_1) F_\Sm(-k_{12}^+) F_\Fm(k_{12}^+) \rint \,.
\end{equation}
If either the $k_1^0$- or $k_2^0$-integration is performed directly, one of the residues involves an eikonal denominator depending on $k_{12}^+$.
This leads to a somewhat unwieldy integral that is difficult to manipulate.
To avoid this problem, it is convenient to rewrite $\IVV{4}$ by first interchanging $k_{1}$ and $k_2$ in the integrand, and then taking half the sum of the two expressions as a new definition 
of $\IVV{4}$.
Hence, we write 
\begin{equation}
\begin{aligned}
    \IVV{4}
    = \frac12 \lint G(k_1) G(k_2) \big[F_\Sm(-k_1) F_\Sm(-k_{12}^+) + F_\Sm(-k_2) F_\Sm(-k_{12}^+)\big] F_\Fm(k_{12}^+) \rint\,.
\label{eq:symmetrization}
\end{aligned}
\end{equation}
The expression in square brackets can be simplified as follows
\begin{equation}
    F_\Sm(-k_1) F_\Sm(-k_{12}^+)
    + F_\Sm(-k_2) F_\Sm(-k_{12}^+)
    = F_\Sm(-k_1) F_\Sm(-k_2) \, .
\label{eq_partial_fractioning}
\end{equation}
Then, applying the transformation $k_{1,2} \mapsto - k_{1,2}$, we obtain
\begin{equation}
    \IVV{4}
    = \frac12 \lint G(k_1) G(k_2) F_\Sm(k_1) F_\Sm(k_2) F_\Fm(-k_{12}^+) \rint \,.
\label{eq_IVV4_rewriting}
\end{equation}
We can now integrate over $k_1^0$ and $k_2^0$ by closing both integration contours in the lower half-plane.
The residues are evaluated at the relevant poles of the gluon propagators $G(k_1)$ and $G(k_2)$ and of the eikonal propagators $F_\Sm(k_1)$ and $F_\Sm(k_2)$.
Before we continue with the integration, we clarify why this procedure does not introduce complications associated with the $\iZero$ prescription.
For definiteness, we focus on the $k_1^0$-integration.

First, consider the relevant pole of the gluon propagator, located at $k_1^0 = \omega_1 - \iZero$.
When we substitute this expression into the eikonal propagator $F_\Fm(-k_{12}^+)$, the induced and existing $\iZero$ terms have the same sign, so no ambiguity arises.
By contrast, when the $k_1^0$-pole is substituted into the eikonal propagator $F_\Sm(k_1)$, the induced and original $\iZero$ terms have \emph{different} signs, so that there is a potential 
for a problem. Luckily, this turns out to be  irrelevant because $2 \Sm \cdot k_1$ cannot vanish for an on-shell $k_1$.
Thus, when computing the contribution from the gluon-propagator pole, we can simply use formulas discussed in the previous section, neglecting the infinitesimal imaginary parts of momentum  $k_{1}$  when evaluating the corresponding residues.

Next, consider the pole of the eikonal propagator $F_\Sm(k_1)$, located at $k_1^0 = \vec k_1 \cdot \vec \Sm - \iZero$.
The imaginary part of $k_1^0$ does not introduce an ambiguity in $F_\Fm(-k_{12}^+)$.
Furthermore, substituting this expression for $k_1^0$ into the gluon propagator yields $1/\big[(\vec \Sm \cdot \vec k_1)^2 - \vec k_1^2 \big]$. The imaginary part of this propagator is irrelevant since, thanks to the fact that the velocity 
of the heavy quark is always smaller than one ($|\vec \Sm | < 1$),  its 
denominator cannot vanish.
%never reach its pole.
The same arguments apply to the integration over the momentum $k_2$.

Combining these observations and using Eqs (\ref{eq_Res_G_LP_rule},
\ref{eq_fermion_pole_UP_LP_plane}),
we obtain
\begin{equation}
\begin{aligned}
    \IVV{4}
    =& -\frac{1}{2} \lint F_\Sm(k_1) F_\Sm(k_2) F_\Fm(k_{12}^+) \rint_{12}
    + \frac{1}{2} \lint G(k_{2,\Sm}) F_\Sm(k_1) F_\Fm(-k_1 - k_{2,\Sm}) \rint_{1\vec2} \\
    & + \frac{1}{8} \big\langle G(k_{1,\Sm}) G(k_{2,\Sm}) F_\Fm(-k_{12,\Sm}^+) \big\rangle_{\vec1 \, \vec2} \,.
\label{eq_IVV4_rewriting_2nd}
\end{aligned}
\end{equation}
We note that the $\langle \mydots \rangle_{\vec1 2}$ and $\langle \mydots \rangle_{1 \vec2}$ contributions have been combined into a single term by relabeling the corresponding momenta.

Our goal is to express the \emph{real part} of the right-hand side of Eq.~(\ref{eq_IVV4_rewriting_2nd}) in terms of the real-virtual and double-real master integrals of Table \ref{tab_master_integrals}. The reason why  only  the real part is needed is that the two-loop amplitude contributes to the cross section through its interference with the tree-level amplitude, $2\operatorname{Re}\!\left[\mathcal M^{(2)}(\mathcal M^{(0)})^*\right]$. 
For the same reason only real parts of all other double-virtual master integrals in Table~\ref{tab_master_integrals} are needed. 

The first term on the \rhs\ of Eq.~(\ref{eq_IVV4_rewriting_2nd}) can be identified with the double-real integral $\IRR{7}$, up to an overall sign.
This identification requires exchanging the momenta $\Fm$ and $\Sm$ in the definition of $\IRR{7}$ and symmetrizing the integral in $k_{1,2}$, following the procedure described above.

To handle the second term on the \rhs\ of \eq\eqref{eq_IVV4_rewriting_2nd}, we consider the real-virtual integral
\begin{equation}
    \IRV{8} = \lint G(k_2) F_\Sm(k_1) F_\Sm(-k_2) F_\Fm(k_{12}^+) \rint_1 \,,
\end{equation}
and integrate over $k_2^0$ by closing the contour in the upper half-plane.
The relevant poles arise from $G(k_2)$ and $F_\Sm(-k_2)$.
Using \eq\eqref{eq_Res_G_UP_rule} for the gluon pole and \eq\eqref{eq_fermion_pole_UP_LP_plane} for the pole of the eikonal propagator, we find
\begin{equation}
    \IRV{8} 
    = - \lint F_\Sm(k_1) F_\Sm(k_2) F_\Fm(k_{12}^-) \rint_{12}
    - \frac{1}{2} \lint G(k_{2,\Sm}) F_\Sm(k_1) F_\Fm(k_1 + k_{2,\Sm}) \rint_{1\vec2} \,.
    \label{eq6.21}
\end{equation}
The first term on the \rhs\ can be dropped because it is purely imaginary, as can be readily shown by applying the replacement $k_{1} \leftrightarrow k_2$.
The second term is nearly identical to the second term on the right-hand side of Eq.~(\ref{eq_IVV4_rewriting_2nd}), except that the arguments of the 
$F_\Fm$ functions have opposite  
signs  in the two expressions.
However, since $F_\Fm(k_1 + k_{2,\Sm}) = - F_\Fm^*(-k_1 - k_{2,\Sm})$ and both $G(k_{2,\Sm})$ and $F_\Sm(k_1)$ are real-valued, the real part of $\IRV{8}$ satisfies the relation
\begin{equation}
    \Re \IRV{8}  
     = \frac{1}{2} \Re\lint G(k_{2,\Sm}) F_\Sm(k_1) F_\Fm(-k_1 - k_{2,\Sm}) \rint_{1\vec2} \,.
\end{equation}

Finally, for the third term on the right-hand side of \eq\eqref{eq_IVV4_rewriting_2nd}, we perform the parity transformation $\vec k_i \to - \vec k_i$, $i=1,2$, and use the following transformation rules for the propagators
\begin{equation}
    F_{\Fm}(-k^+_{12,\Sm}) \to 
    - F_{\Fm}^*(-k^+_{12,\Sm}) \,,
    \qquad G(k_{i,\Sm}) \to G(k_{i,\Sm}) \,,
    \quad  i = 1,2 \,.
\end{equation}
Since $G(k_{i,\Sm})$ is real, these transformation rules imply that the third term in Eq.~(\ref{eq_IVV4_rewriting_2nd}) is purely imaginary and therefore can be dropped.

Combining all these results, we obtain the desired expression for the real part of the double-virtual integral $\IVV{4}$  in terms of real-virtual and double-real integrals
\begin{equation}
     \Re \IVV{4} = \Re \IRV{8} - \IRR{7} \,.
\label{eq_IVV4_IRV8_IRR7_cancellation_gl_res}
\end{equation}
This equation is illustrated in Figure \ref{fig:IVV4Rel}, which also explains why it is natural to expect a connection between the three master integrals.
This expectation follows from the 
fact that  a dispersion relation for the relevant two-point function together 
with  
analyticity arguments can be used to assert that its  full imaginary part should be  free of $\order{\mg}$ contributions.  Since the imaginary 
part can be computed from the 
individual cuts of the two-point function that obviously  contain ${\cal O}(m_g) $ terms, relations between contributions of different cuts must exist. Eq.~(\ref{eq_IVV4_IRV8_IRR7_cancellation_gl_res}) provides an example of such a relation.  Furthermore,  such relations   can be used to motivate choices of the real-virtual integrals that may be linked  to particular double-virtual master integrals.
\begin{figure}[t]
    \centering 
    \begin{align*}
        \Re\left[\vcenter{\hbox{\FigIVVfour[0.35]}}\right]
        = \Re\left[\vcenter{\hbox{\FigIRVeight[0.35]}}\right]
        - \vcenter{\hbox{\FigIRRseven[0.35]}}
    \end{align*}
    \vspace{-5mm}
    \caption{Graphical representation of the identity in \eq\eqref{eq_IVV4_IRV8_IRR7_cancellation_gl_res}.}
    \label{fig:IVV4Rel}
\end{figure}

%================================================
% IVV{1}
%================================================
\subsection{Integral with three gluon and three eikonal propagators}
\label{sect:ivv1}
We continue  the discussion with the double-virtual integral $\IVV{1}$.
Our goal is to relate its real part to the real-virtual and double-real integrals.
Although we follow the steps described in the previous section, this example involves a new element, 
mentioned after \eq\eqref{eq_Res_G_LP_rule}, i.e.\ the need to retain
the infinitesimal imaginary part of  the loop momentum that arises after taking 
the first residue in the subsequent calculation.
The integral reads (cf.\ Table~\ref{tab_master_integrals})
\begin{equation}
    \IVV{1} 
    = \vcenter{\hbox{\FigIVVone[0.35]}}
    = \lint G(k_1) G(k_2) G(k_{12}^-) F_\Fm(k_1) F_\Sm(-k_1) F_\Sm(-k_2) \rint \,. % \dmtc
    \label{eq6.25}
\end{equation}
We first integrate over $k_2^0$ closing the contour in the lower half-plane (thus avoiding the pole of $F_\Sm(-k_2)$).
The poles of $G(k_2)$ and $G(k_{12}^-)$ give rise to two contributions
\begin{equation}
    \IVV{1}= \IVV{\rm 1A} + \IVV{\rm 1B} \,, % \dmtc
    \label{eq6.26}
\end{equation}
where $\IVV{\rm 1A}$ corresponds to the residue of $G(k_2)$ and $\IVV{\rm 1B}$ to the residue of $G(k_{12}^-)$.

To illustrate the subtleties that arise in evaluating these integrals, we first consider $\IVV{\rm 1A}$.
When we take the pole of $G(k_2)$, all the remaining propagators in Eq.~(\ref{eq6.25}) must be evaluated at the on-shell value of $k_2$.
Consider the propagator $G(k_{12}^-)$.
At the pole, it evaluates to
\begin{equation}
\begin{aligned}
    G(k_1 - k_2^\rmc)\left|\substack{(k_2^\rmc)^2 \,=\, \mg^2 - \iZero \\ k_{2,0}^\rmc \,=\, \om_2 - \iZero \hspace{1.7mm} } \right.
    & = \frac{1}{(k_{12}^-)^2 - \mg^2 + \iZero\,\sign(k_1^0)}
    \bigg|_{k_2^2 \,=\, \mg^2} \\
    & = \Big[G(k_{12}^-) + 2\pi i \, \theta(-k_1^0) \delta[( k_{12}^-)^2 - \mg^2] \Big ]_{k_2^2 \,=\, \mg^2} \,, % \dmtc
\label{eq_G_k12_k2_pole_relation}
\end{aligned}
\end{equation}
where $k_2^\rmc = (\om_2 - \iZero, \veck_2)$ and $k_2 = (\om_2, \vec k_2)$.
The delta-function term in \eq\eqref{eq_G_k12_k2_pole_relation} accounts for the change in the $\iZero$ prescription and must be included in $\IVV{\rm 1A}$ together with the contribution on the \rhs\ of \eq\eqref{eq_Res_G_LP_rule}.
We note that $G(k_{12}^-)$ is the only propagator in $\IVV{1}$ that requires care after taking the positive-energy residue of $G(k_2)$.
We therefore write
\begin{equation}
    \IVV{\rm 1A} = \IVV{\rm 1A,1} + \IVV{\rm 1A,2} \,, % \dmtc
\end{equation}
with
\begin{equation}
\begin{aligned}
    & \IVV{\rm 1A,1} = - \lint  G(k_1) G(k_{12}^-) F_\Fm(k_1) F_\Sm(-k_1) F_\Sm(-k_2) \rint_2 \,, \\
    & \IVV{\rm 1A,2} =  - 2\pi i \, \lint \theta(-k_1^0) \delta[ (k_{12}^-)^2 - \mg^2] \, G(k_1) F_\Fm(k_1) F_\Sm(-k_1) F_\Sm(-k_2) \rint_2 \,. % \dmtc
\label{eq_IVV_1A1_1A2}
\end{aligned}
\end{equation}

The second contribution on the 
right-hand side of Eq.~(\ref{eq6.26}), $\IVV{\rm 1B}$, originates from the 
$k_2$-pole of $G(k_{12}^-)$ located in the lower half-plane.
The simplest way to evaluate this contribution is to shift $k_2 \to k_1 + k_2$ and then repeat the above procedure.
This gives
\begin{equation}
    \IVV{\rm 1B}= \IVV{\rm 1B,1} + \IVV{\rm 1B,2} \,, % \dmtc
\end{equation}
where
\begin{equation}
\begin{aligned}
    & \IVV{\rm 1B,1} = - \lint  G(k_1) G(k_{12}^+)F_\Fm(k_1) F_\Sm(-k_1) F_\Sm(-k_{12}^+) \rint_2 \,, \\
    & \IVV{\rm 1B,2} = -2\pi i \, \lint \theta(k_1^0) \delta[ (k_{12}^+)^2 - \mg^2] \, G(k_1) F_\Fm(k_1) F_\Sm(-k_1) F_\Sm(-k_{12}^+) \rint_2 \,. % \dmtc
\label{eq_IVV_1B1_1B2}
\end{aligned}
\end{equation}

To proceed, we first combine the integrals $\IVV{\rm 1A,2}$ and $\IVV{\rm 1B,2}$.
In $\IVV{\rm 1A,2}$, we shift $k_1 \to k_1 + k_2$ and find
\begin{equation}
    \IVV{\rm 1A,2} = -2\pi i \, \lint \theta(-k_{12,0}^+) \delta( k_1^2 - \mg^2) \, G(k_{12}^+) F_\Fm(k_{12}^+) F_\Sm(-k_{12}^+) F_\Sm(-k_2) \rint_2 \,. % \dmtc
\end{equation}
The Heaviside function selects the negative-energy contribution $k_1^0 = -\omega_1$, while the positive-energy contribution vanishes.
Expressing the result in terms of the positive-energy on-shell momenta, we obtain
\begin{equation}
    \IVV{\rm 1A,2} = - \lint \theta(\omega_1 - \omega_2) \, G(k_{12}^-) F_\Fm(-k_{12}^-) F_\Sm(k_{12}^-) F_\Sm(-k_2) \rint_{12} \,. % \dmtc
\label{eq_IVV_1A2_fin}
\end{equation}
To simplify $\IVV{\rm 1B,2}$, we shift $k_1 \to k_1 - k_2$ and find
\begin{equation}
    \IVV{\rm 1B,2} = -2\pi i \, \lint \theta(k_{12,0}^-) \delta (k_1^2 - \mg^2) \, G(k_{12}^-) F_\Fm(k_{12}^-) F_\Sm(-k_{12}^-) F_\Sm(-k_1) \rint_2 \,. % \dmtc
\end{equation}
In this case, the Heaviside function selects the positive-energy contribution $k_1^0 = \omega_1$, yielding
\begin{equation}
    \IVV{\rm 1B,2} = - \lint \theta(\omega_1 - \omega_2) \, G(k_{12}^-) F_\Fm(k_{12}^-) F_\Sm(-k_{12}^-) F_\Sm(-k_1) \rint_{12} \,. % \dmtc
\label{eq_IVV_1B2_fin}
\end{equation}
We then exchange the labels $1$ and $2$ in \eq\eqref{eq_IVV_1A2_fin} and add the resulting expression to \eq\eqref{eq_IVV_1B2_fin}.
Combining the  two Heaviside functions,
we obtain
\begin{equation}
    \IVV{\rm 1A,2} + \IVV{\rm 1B,2}
    = \lint G(k_{12}^-) F_\Fm(k_{12}^-) F_\Sm(-k_{12}^-) F_\Sm(k_1) \rint_{12} \,, % \dmtc
\label{eq_IVV_1A2_1B2_sum}
\end{equation}
where we have used $F_\Sm(-k_1) = -F_\Sm(k_1)$ for on-shell $k_1$.
The result for the two integrals in Eq.~(\ref{eq_IVV_1A2_1B2_sum}) will be combined below with the remaining contributions from $\IVV{\rm 1A,1}$ and $\IVV{\rm 1B,1}$.

We begin with the integral $\IVV{\rm 1B,1}$ defined in \eq\eqref{eq_IVV_1B1_1B2}.
Applying the partial-fraction identity  in \eq\eqref{eq_partial_fractioning}, we obtain
\begin{equation}
\begin{aligned}
    \IVV{\rm 1B,1} 
    = & - \lint G(k_1) G(k_{12}^+)F_\Fm(k_1) F_\Sm(-k_1) F_\Sm(-k_2)\rint_2 \\
    & + \lint G(k_1) G(k_{12}^+) F_\Fm(k_1) F_\Sm(-k_2) F_\Sm(-k_{12}^+) \rint_2 \,.
\end{aligned} % \dmtc
\end{equation}
After the replacement $k_1 \mapsto - k_1$ in the first term, the expression for $\IVV{\rm 1B,1}$ can be combined with $\IVV{\rm 1A,1}$ and written as the sum of two contributions
\begin{equation}
    \IVV{\rm 1A,1} + \IVV{\rm 1B,1}
    = \IVV{\rm 1C} + \IVV{\rm 1D} \,, % \dmtc
\end{equation}
where
\begin{equation}
\begin{aligned}
    \IVV{\rm 1C} & = \lint G(k_1) G(k_{12}^+) F_\Fm(k_1) F_\Sm(-k_2) F_\Sm(-k_{12}^+) \rint_2 \,, \\
    \IVV{\rm 1D} & = - \lint G(k_1) G(k_{12}^-)\big[F_\Fm(k_1) F_\Sm(-k_1) + F_\Fm(-k_1) F_\Sm(k_1)\big]F_\Sm(-k_2)\rint_2 \,.
\end{aligned} 
\label{eq6.39} % \dmtc
\end{equation}

To simplify $\IVV{\rm 1C}$, we replace $k_1 \mapsto -k_1$ and use \eq\eqref{eq:GF-identities} together with the fact that $F_\Sm(-k_2)$ is real-valued to write
\begin{equation}
    \IVV{\rm 1C} = - \lint G^*(k_1) G^*(k_{12}^-)F_\Fm(k_1)
    F_\Sm(-k_2) F_\Sm(-k_{12}^-) \rint^*_2 \,. % \dmtc
\end{equation}
We then replace the complex-conjugated gluon propagators with the regular ones using \eq\eqref{eq:GF-identities} and find
\begin{align}
    \IVV{\rm 1C} 
    = & - \lint G(k_1) G(k_{12}^-) F_\Fm(k_1)
    F_\Sm(-k_2) F_\Sm(-k_{12}^-) \rint^*_2 \notag \\
    & + 2\pi i\, \lint \delta(k_1^2-\mg^2) \, G(k_{12}^-)F_\Fm(k_1) F_\Sm(-k_{12}^-)
    F_\Sm(-k_2) \rint^*_2 
    \label{eq_I1C_expression} \\
    & + 2\pi i\, \lint \delta[(k^-_{12})^2-\mg^2]\, G(k_{1}) F_\Fm(k_1) F_\Sm(-k_{12}^-) F_\Sm(-k_2) \rint^*_2 \,. \notag  % \dmtc
\end{align}
We note that the product of two delta functions arising from the two gluon propagators requires
\begin{equation}
    k_1^2 = \mg^2 \,, 
    \qquad 
    (k_1 - k_2)^2 = \mg^2 \,.
\end{equation}
These equations have no support for the real-valued on-shell $k_{1,2}$; for this reason,  the corresponding contribution has been dropped from Eq.~(\ref{eq_I1C_expression}).

The first term in Eq.~(\ref{eq_I1C_expression}) can be related to one of the real-virtual integrals.
The simplification of the remaining two terms requires the evaluation of the other contributions.
To simplify $\IVV{\rm 1D}$ in Eq.~(\ref{eq6.39}), we use $F_\Fm(k_1) = - F_\Fm^*(-k_1)$ together with the reality of $F_\Sm(-k_2)$, and find 
\begin{align}
    \IVV{\rm 1D} 
    =& - \lint G(k_1) G(k_{12}^-)\big[F_\Fm(k_1) F_\Sm(-k_1) + F_\Fm^*(k_1) F_\Sm^*(-k_1)\big] F_\Sm(-k_2) \rint_2 \notag \\
    =& - \Big[\lint G(k_1) G(k_{12}^-) F_\Fm(k_1) F_\Sm(-k_1)F_\Sm(-k_2) \rint_2 - {\rm c.c.} \Big] \notag \\
    & - 2\pi i \, \lint \delta(k_1^2 - \mg^2) \, G(k_{12}^-)  F_\Fm(k_1) F_\Sm(-k_1) F_\Sm(-k_2) \rint_2^* 
    \label{eq6.43} \\
    & - 2\pi i \, \lint \delta[(k_{12}^-)^2 - \mg^2] \, G(k_1)  F_\Fm(k_1) F_\Sm(-k_1) F_\Sm(-k_2) \rint_2^* \,. \notag % \dmtc
\end{align}
As in the discussion of the $\IVV{\rm 1C}$ contribution, the term involving two delta functions has been dropped because it has no support.
The combination of the integral and its complex conjugate in the second line of Eq.~(\ref{eq6.43}) is purely imaginary.
Hence, we write
\begin{equation}
\begin{aligned}
    \IVV{\rm 1D} 
    = & - 2\pi i \, \lint \delta(k_1^2 - \mg^2) \, G(k_{12}^-)  F_\Fm(k_1) F_\Sm(-k_1) F_\Sm(-k_2) \rint_2^* \\
    & - 2\pi i \, \lint \delta[(k_{12}^-)^2 - \mg^2] \, G(k_1)  F_\Fm(k_1) F_\Sm(-k_1) F_\Sm(-k_2) \rint_2^* % \dmtc
    + \mydots \,,
\label{eq_I1D_expression}
\end{aligned}
\end{equation}
where the ellipsis denotes a purely imaginary contribution whose explicit form is irrelevant for our purposes.

%%%%%%%%%%%%%%%%%%%%%%%%%%%%%%%%%%%%%%%%%%%%%%%%%%%%%%%%
% FIGURE 3
%%%%%%%%%%%%%%%%%%%%%%%%%%%%%%%%%%%%%%%%%%%%%%%%%%%%%%%%
\begin{figure}[t]
    \centering 
    \begin{align*}
        \Re\left[\vcenter{\hbox{\FigIVVone[0.35]}}\right]
        =&\, \Re\left[\vcenter{\hbox{\FigIRVone[0.35]}}\right]
        + \Re\left[\vcenter{\hbox{\FigIRVtwo[0.35]}}\right] \\
        & - \hspace{4.35mm} \vcenter{\hbox{\FigIRRone[0.35]}}
        \hspace{5mm}-
        \hspace{5.3mm} \vcenter{\hbox{\FigIRRtwo[0.35]}}
    \end{align*}
    \vspace{-5mm}
    \caption{Graphical representation of the identity in \eq\eqref{eq_Ivv1_final}.}
    \label{fig:IVV1Rel}
\end{figure}
Combining Eqs~\eqref{eq_IVV_1A2_1B2_sum}, \eqref{eq_I1C_expression}, and \eqref{eq_I1D_expression}, we obtain
\begin{align}
    \IVV{1} 
    & = \IVV{\rm 1A,2} + \IVV{\rm 1B,2} + \IVV{\rm 1C} + \IVV{\rm 1D}
    = - \lint G(k_1) G(k_{12}^-) F_\Fm(k_1) F_\Sm(-k_{12}^-) F_\Sm(-k_2) \rint_2^* \notag \\
    & - \lint G(k_{12}^+) F_\Fm(-k_1) F_\Sm(k_{12}^+) F_\Sm(k_1) \rint_{12}^*
    - \lint G(k_{12}^+) F_\Fm(k^+_{12})F_\Sm(-k_1) F_\Sm(-k_{12}^+) 
    \rint_{12}^* \hspace{0.5cm}
    \label{eq6.45} \\
    & - \lint G(k_{12}^-) F_\Fm(k_1) F_\Sm(-k_{12}^-) F_\Sm(-k_1) \rint_{12}^*
    + \mydots \,. \notag % \dmtc
\end{align} 
The last term in the above equation  involves integration over two on-shell gluon momenta.
However, this integral cannot be related to any of the double-real integrals because the momentum $k_1 - k_2$ appears in one of the eikonal propagators.
This difference in sign can be interpreted as the emission of a gluon with momentum $k_1$ and the absorption of a gluon with momentum $k_2$.
To describe the latter, we consider the following real-virtual integral
\begin{equation}
    \calI_{\rm RV} = \lint G(k_2) G(k_{12}^-) F_\Fm(k_1) F_\Sm(k_{12}^-) F_\Sm(k_1) \rint_1 \,, % \dmtc
\end{equation}
and integrate over $k_2^0$ by closing the contour in the lower half-plane.
We find
\begin{equation}
    \calI_{\rm RV} 
    = - \lint G(k_{12}^-) F_\Fm(k_1) F_\Sm(k_{12}^-) F_\Sm(k_1) \rint_{12}
    - \lint G(k_{12}^+) F_\Fm(k_1) F_\Sm(-k_2) F_\Sm(k_1) \rint_{12} \,, % \dmtc
\end{equation}
so that
\begin{equation}
    \lint G(k_{12}^-) F_\Fm(k_1) F_\Sm(k_{12}^-) F_\Sm(k_1) \rint_{12}
    = - \calI_{\rm RV} 
    - \lint G(k_{12}^+) F_\Fm(k_1) F_\Sm(-k_2) F_\Sm(k_1) \rint_{12} \,.
    \label{eq4.48} % \dmtc
\end{equation}
Since the reality of $G(k_{12}^-)$, $F_\Fm(k_1)$ and $F_\Sm(-k_1)$ ensures the validity of the following relation
\begin{equation}
    \lint G(k_{12}^-) F_\Fm(k_1) F_\Sm(-k_{12}^-) F_\Sm(-k_1) \rint_{12}^* =
    \lint G(k_{12}^-) F_\Fm(k_1) F_\Sm(k_{12}^-) F_\Sm(k_1) \rint_{12} \,, % \dmtc
\end{equation}
we can use Eq.~(\ref{eq4.48}) to simplify $\IVV{1}$ in Eq.~(\ref{eq6.45}).
Combining two of the double-real terms using the partial-fraction identity in \eq\eqref{eq_partial_fractioning}, we obtain
\begin{align}
    \IVV{1} 
    = & - \lint G(k_1) G(k_{12}^-) F_\Fm(k_1) F_\Sm(-k_{12}^-) F_\Sm(-k_2) \rint_2^* \notag \\
    & + \lint G(k_2) G(k_{12}^-) F_\Fm(k_1) F_\Sm(k_{12}^-) F_\Sm(k_1) \rint_1
    - \lint G(k_{12}^+) F_\Fm(k^+_{12})F_\Sm(k_1) F_\Sm(k_{12}^+) 
    \rint_{12} \\
    & - \lint G(k_{12}^+) F_\Fm(k_1)F_\Sm(k_{12}^+) F_\Sm(k_2) 
    \rint_{12}
    + \mydots \,. \notag %  \dmtc
\end{align}
The four terms on the right-hand side can be identified with the master integrals listed in Table~\ref{tab_master_integrals}.
Taking the real part of $\IVV{1}$, we arrive at the desired result
\begin{equation}
    \Re \IVV{1} = \Re\IRV{1} + \Re\IRV{2} - \IRR{1} - \IRR{2} \,, % \dmtc
\label{eq_Ivv1_final}
\end{equation}
which is graphically represented in Figure \ref{fig:IVV1Rel}.

%================================================
% RESULTS of IVV MAPPING
%================================================
\subsection{Results for the double-virtual integrals}
\label{subsec_results_Ivv}
The results for all other double-virtual integrals can be obtained in a similar way.
Below, we list the results for the ten soft-soft double-virtual integrals listed in Table~\ref{tab_master_integrals},%
\footnote{We note that 
real-emission integrals where 
a massive gluon splits into a pair of massless quarks   are poorly defined and require 
resummation of self-energy 
insertions turning the gluon  
propagator into a Breit-Wigner 
function. The pole of this function corresponds to the 
production of a single on-shell gluon that we have treated at NLO. The ``off-shell'' contribution then cancels with the double-virtual and real-virtual integrals as shown in Eq.~(\ref{eq:NNLO:mappingVVRVRR}). 
While this subtlety is an unfortunate consequence of incorporating massless quarks into our model, we believe that it is not essential for
the discussion of the linear sensitivity of $t \bar t $ production.}
\begin{align} 
    \Re \IVV{1} & = \Re \IRV{1} + \Re \IRV{2}  - \IRR{1} - \IRR{2} \,, \notag \\
    %------------------------------------------------
    \Re \IVV{2} & = \Re \IRV{9} \,, \notag \\
    %------------------------------------------------
    \Re \IVV{3} & = - \Re \IRV{7} + \Re \IRV{3} + \IRR{8} \,, \notag \\
    %------------------------------------------------
    \Re \IVV{4} & = \Re \IRV{8} - \IRR{7} \,, \notag \\
    %------------------------------------------------
    \Re \IVV{5} & = - \Re \IRV{4} + \Re \IRV{5} - \Re \IRV{6} + \IRR{3} - \IRR{5} + \IRR{6} \,, 
    \label{eq:NNLO:mappingVVRVRR} \\
    %------------------------------------------------
    \Re \IVV{6} & = - \bigg(\frac{\Fm^2}{\Fm \cdot \Sm}\bigg) \Big[\Re \IRV{13} + \IRR{11} \Big]
    - \bigg(\frac{3-4 \epsilon}{8 \mg^2 (\Fm \cdot \Sm)}\bigg) \Big[\Re \IRV{15} + \IRR{12} \Big] 
    + \Re \IRV{14} \,, \hspace{10mm}\notag \\
    %------------------------------------------------
    \Re \IVV{7} & = \IR{2} \Re \IV{4} - \IRR{14} \,, \notag \\
    %------------------------------------------------
    \Re\IVV{8} & = - 2\Re\IRV{11} + \Re\IRV{12} + 2 \IRR{9} - \IRR{10} \,, \notag \\
    %------------------------------------------------
    \Re\IVV{9} & = - 2\Re\IRV{15} - \IRR{12} \,, \notag \\
    %------------------------------------------------
    \Re\IVV{10} & = - \Re\IRV{16} \,. \notag 
\end{align}
We note that the $\ep$-dependent coefficients of the real-virtual integrals in the equation for $\IVV{6}$ arise because, after the residues have been evaluated, an additional IBP reduction is performed to express the resulting real-virtual integrals in terms of the master integrals.

\section{Renormalization and mass shift of tree-level and one-loop amplitudes}
\label{sect:7}

In this section, we discuss the renormalization and the mass shift of 
the tree-level and one-loop amplitudes.
The procedure largely follows the NLO discussion presented in \Sec\ref{sec:4}.
However, since  NNLO calculations are naturally more involved, we provide additional details below.

As in the NLO case, we renormalize the mass and the wave function of the top quark and of the gluon in the on-shell scheme.
The renormalization constants required for this procedure have been computed in Ref.~\cite{Fontes:2026ihz}.

We will consider renormalization schemes where 
the coupling renormalization constant $Z_g^{(1)}$ in
Eq.~(\ref{eq:DefCouplingAndSDMass})
does not depend on powers of the gluon mass $m_g$. 
Since the leading-order cross section for $t \bar t $ production is independent of the coupling constant $\gs$, and since the linear $m_g$-contribution to  the NLO 
cross section cancels once the appropriate mass definition is used, one expects the dependence on $Z_g^{(1)}$ in the ${\cal O}(m_g)$ part of the NNLO cross section to disappear without the need to specify its explicit form.

As at NLO, the cancellation of ${\cal O}(m_g)$ contributions to the cross section can occur only if it is expressed in terms of a short-distance mass.
The precise definition of this mass is, however, not needed.
We introduce the relation between the pole mass and a 
short-distance mass $\mtt$ by writing
\be
\mt = \frac{Z_{\mtt}}{Z_{\mt} } \, \mtt \, ,
\ee
where $Z_{\mtt, \mt}$ are the renormalization constants relating the bare top-quark mass to $\mtt$ or $m_t$, respectively. By construction, the perturbative expansion of $Z_{\mtt}$ contains no ${\cal O}(m_g)$ terms and remains arbitrary otherwise.
We note that only the one-loop contribution to that renormalization constant, $Z_{\mtt}^{(1)}$, is required.
We  keep this renormalization constant unspecified, and  expect all ${\cal O}(m_g)$ terms involving $ Z^{(1)}_{\mtt}$ to cancel independently of the remaining contributions.
We use the pole-mass renormalization constant $Z_{\mt}$ computed in Ref.~\cite{Fontes:2026ihz}.\footnote{We note that in Ref.~\cite{Fontes:2026ihz} the analytical expression for $Z_{\mt}$ was computed. We   rewrite it back in terms of master integrals, which are a subset of the master integrals in Appendix~\ref{appendix_sec_list_MIs}. We emphasize that  
$Z_{\mt}$  contains ${\cal O}(m_g)$ terms.}

While  the application of the renormalization procedure to one-loop virtual amplitudes (as well as to Born $t \bar t g$ amplitudes) is straightforward, it must be organized carefully.
For example, it involves mass counterterms that cancel the ${\cal O}(m_g^{-1})$ terms arising from self-energy insertions on nearly on-shell quark lines.
Furthermore, we account for the change in the one-loop contribution induced by replacing the pole mass with the short-distance mass in two steps. We  start by expressing the master integrals and their coefficients in terms of the heavy-quark velocities, and then  compute changes in these velocities caused by the mass shift.
Since the one-loop virtual and tree-level real-emission amplitudes for $t \bar t g$ contain both $\order{\mg^0}$ and $\order{\mg}$ terms, the mass shift must be taken into account in both contributions.
\section{The cancellation of ${\cal O}(\mg)$ contributions to heavy-quark production cross section at NNLO}
\label{sect:8}

In this section, we discuss the cancellation of ${\cal O}(\mg)$ 
contributions to the heavy-quark production cross section at NNLO. 
If this cancellation does happen, it should occur   independently for several
groups of terms. For example, while contributions from the renormalization constants and the mass shift involve two-loop on-shell self-energy integrals,
corrections to the vertex generate very different integrals, which depend 
on the  scalar product of four-momenta of heavy quarks, $\pt \cdot \ptb$.  It is 
obvious that these contributions have 
to cancel independently, if at all. 
In total, we have identified five different classes 
of the ${\cal O}(\mg)$ contributions to the  NNLO cross section that vanish independently of each other. We discuss them below.\footnote{There are other standard 
handles that allow one to split the full NNLO correction into different parts, e.g.\ the dependence on the color factors, the number of massless  quarks or the dependence on the Higgs boson mass.  
We have used them  to debug the calculation,  but they do not play 
an important  role in establishing full cancellation.} \\

\noindent \textbf{Class A} includes soft-soft double-virtual, double-real and real-virtual integrals that depend on the scalar product $p_t \cdot p_{\bar t} $. 
Defining $y=\tilde\beta^2$ and
$\Lambda\equiv 1-2\epsilon-2y(1-\epsilon)$, we find
\begin{align}
    &\; \mathcal{T}_{\mg}\left[
    \frac{\mathrm{d}\sigma_{t\bar t}^{(2)}}{\mathrm{d}\Omega_t}
    \right]_{\,\mathclap{\rmA}}
    = \frac{{\cal N}^{\,2}\,\sigma_0}{2\sqrt{s}\tilde\beta^{1+2\epsilon}}
    (3-y-2\epsilon) \bigg\{8\Cf^2\Bigl[
    (1-y)^2(1+y)\Big(I_{\rm RR}^{(8)}+\Re I_{\rm RV}^{(3)}
    -\Re I_{\rm VV}^{(3)} \notag \\
    & -\Re I_{\rm RV}^{(7)}\Big)
    +(1-y)^3 \Big(I_{\rm RR}^{(7)}-\Re I_{\rm RV}^{(8)}
    +\Re I_{\rm VV}^{(4)} \Big)\Bigr]
    -\frac{4\Ca\Cf y}{(1+y)(1-2\epsilon)}
    \bigg[4(y-1)\Lambda \Big(I_{\rm RR}^{(11)} \notag \\[.2em]
    & +\Re I_{\rm RV}^{(13)} \Big)
    +(1+y)\Bigl[(1-2\epsilon)(3+6y-y^2)
    \Big(I_{\rm RR}^{(7)}-\Re I_{\rm RV}^{(8)}+\Re I_{\rm VV}^{(4)} \Big) \\[.2em]
    & +(1-2\epsilon)(1+4y-y^2)
    \Big(\Re I_{\rm RV}^{(7)}-\Re I_{\rm RV}^{(3)}
    +\Re I_{\rm VV}^{(3)}-I_{\rm RR}^{(8)} \Big) 
    -6\mg^2(1-2\epsilon)(1+y) \notag \\[.2em]
    & \times \Big(\Re I_{\rm RV}^{(1)}+\Re I_{\rm RV}^{(2)}
    -\Re I_{\rm VV}^{(1)}-I_{\rm RR}^{(1)}-I_{\rm RR}^{(2)} \Big)
    +4\Lambda \Big(\Re I_{\rm RV}^{(14)}-\Re I_{\rm VV}^{(6)} \Big) 
    -2\mg^2(1-9y \notag \\[.2em]
    & -4\epsilon+12y\epsilon) \Big(\Re I_{\rm RV}^{(4)}-\Re I_{\rm RV}^{(5)}
    +\Re I_{\rm RV}^{(6)}+\Re I_{\rm VV}^{(5)}
    -I_{\rm RR}^{(3)}+I_{\rm RR}^{(5)}-I_{\rm RR}^{(6)} \Big)
    \Bigr]\bigg]\bigg\} \,. \notag
\end{align}
We discussed  in Section~\ref{sect:MI} how to express all double-virtual soft 
integrals through the real-virtual and double-real ones. Using the results shown in 
Eq.~(\ref{eq:NNLO:mappingVVRVRR}) to remove the double-virtual integrals from  the ${\cal O}(\mg)$ contribution to ${\rm d} \sigma^{(2)}_{t \bar t}$, we find  that  $\mathcal{T}_{\mg} \big[\rmd \sigma_{t\bart}^{(2)}/{\rm d} \Omega_t\big]_\rmA \propto \IRR{12}$, which does not depend on $\pt \cdot \ptb$ and, therefore, belongs to class B, which is discussed below. \\

\noindent \textbf{Class B} includes genuine two-loop integrals that depend 
on $p_t^2 = p_{\bar t}^2 = m_t^2$ and 
not on $p_t \cdot p_{\bar t}$. 
Such integrals arise from the two-loop 
wave-function renormalization, from the two-loop mass shift of the LO cross section, and from 
${\cal O}(\mg)$ soft-soft double-virtual, real-virtual and double-real corrections to $t \bar t$ production. Defining \(\delta=1-2\epsilon\), we find\footnote{We note that Eq.~\eqref{eq:ClassB} includes the leftover from class~A, as discussed above.}
\begin{align} 
    &\, \mathcal{T}_{\mg}\left[
    \frac{\mathrm{d}\sigma_{t\bar t}^{(2)}}{\mathrm{d}\Omega_t}
    \right]_{\,\mathclap{\rmB}}
    = \frac{{\cal N}^{\,2}\sigma_0 \Cf(3-y-2\epsilon)}
    {16\sqrt{s}\,\mg^2\tilde\beta^{1+2\epsilon}}\biggl\{
    8(1+2\delta) \Bigl[\Cf(1+y)^2
    \Big(I_{\rm RR}^{(12)}+2\Re I_{\rm RV}^{(15)}+\Re I_{\rm VV}^{(9)} \Big) \notag \\[.2em]
    & -2\mg^2 \TR \nf \,y(1+\delta)
    \Big(I_{\rm RR}^{(14)}+\Re I_{\rm VV}^{(7)} \Big)\Bigr]
    +\Ca y\biggl[\mg^2\bigl[23+6\delta(21+4\delta)\bigr]
    \Big(I_{\rm RR}^{(10)}+\Re I_{\rm VV}^{(8)} \Big) \notag \\[.2em]
    & -\frac{1+2\delta}{\delta}\bigg(
    \big[8+\delta(25+4y+20\delta)\big]
    \Big(I_{\rm RR}^{(12)}+\Re I_{\rm VV}^{(9)}\Big)
    +\frac{8\bigl(2+4y+\delta(y^2-5)\bigr)}{1+y}
    \,\Re I_{\rm RV}^{(15)}\bigg) \notag \\[.2em]
    & -32\mg^2(1+3\delta)
    \Big(2I_{\rm RR}^{(9)}-2\Re I_{\rm RV}^{(11)}+\Re I_{\rm RV}^{(12)}\Big)
    \biggr]\biggr\} \,. \label{eq:ClassB}
\end{align}
Similarly to class~A, upon inserting the mappings in Eq.~\eqref{eq:NNLO:mappingVVRVRR}, we observe that all genuine two-loop 
integrals in this class cancel  
and only products of one-loop integrals remain. The latter are combined with other similar integrals into class C.
We emphasize that, in order to obtain this result, the on-shell scheme renormalization constants must be rewritten in terms of master integrals. \\

\noindent \textbf{Class C} contains contributions represented by products of one-loop soft integrals. These contributions arise from the mass-shift in one-loop soft contributions (both real and virtual), from two-loop contributions and from the leftovers in Class B.
We find that they  cancel. \\

\noindent \textbf{Class D} comprises hard-soft contributions that arise from the double-virtual and real-virtual corrections, the renormalization and mass redefinition of the hard and soft virtual and  real contributions at NLO, as well as the mass shift of the tree-level cross section. 
We find that these contributions cancel as well. \\

\noindent \textbf{Class E} contains one-loop soft integrals 
multiplied by the one-loop strong-coupling and short-distance mass renormalization constants.
Such terms originate from the  renormalization and mass shift in LO and NLO contributions.
Upon relating the integrals to each other, we find that the dependence on these constants cancels as well. We stress that the cancellation occurs with both $Z_g^{(1)}$ and $Z_{\mtt}^{(1)}$ left unspecified. \\

The five classes discussed above cover 
all $\order{\mg}$ contributions to the $t \bar t$ production cross section at NNLO. Since each of these classes vanishes, 
we conclude that the 
$t \bar t $ production 
cross section has no ${\cal O}(\mg)$ terms through NNLO provided that it is 
expressed in terms of a short-distance top-quark mass 
\be
\mathcal{T}_{\mg} \Bigg[\frac{ {\rm d} \sigma_{t\bart} }{{\rm d} \Omega_t}\Bigg] = \order{\mg \as^3} \, .
\ee
To avoid confusion, we emphasize that this result has been  obtained in a non-Abelian ${\rm SU}(2)$ gauge theory with spontaneous symmetry breaking, described 
in Section~\ref{sec:model}, and \emph{not} in QCD.

\section{Conclusions}
\label{sect:9}

In this paper, we have studied the production of a heavy quark and an antiquark from a neutral source in a theory with a spontaneously broken non-Abelian $\SU{2}$ gauge group.
In this theory, three gauge bosons acquire equal masses, which we denote by $\mg$.
Assuming that $\mg$ is the smallest dimensionful parameter in the theory, we can investigate the dependence of the $t \bar t$ production cross section on $\mg$ in the limit $\mg \ll \sqrt{s} \sim \mt$.

The leading dependence on $\mg$ is logarithmic; it appears separately in virtual and real-emission contributions, as well as in the heavy-quark on-shell renormalization constants.
However, the KLN theorem \cite{Kinoshita:1962ur,Lee:1964is}
ensures that all $\order{\log \mg}$ contributions cancel in infrared-safe observables, leaving behind an $m_g$-independent contribution.  Generalizing from our model, this is the standard contribution that one computes  in   perturbative  QCD to describe hard processes at colliders.

The first subleading correction to this result is \emph{linear} in the gluon mass, i.e.\ $\order{\mg}$.
In the context of renormalon-based analyses, such dependence is used to estimate the sensitivity of a particular process or observable to linear ${\cal O}(\Lambda_{\rm QCD})$ power corrections.
Unfortunately, most of the existing renormalon-based analyses of collider processes and observables are restricted to the Abelian subsector of QCD, and a generalization to cases that are sensitive to non-Abelian interactions is desirable.
Since the toy model that we use in this paper is a renormalizable quantum field theory, the linear sensitivity of the $t \bar t$ production cross section to the gluon mass $\mg$ can be studied rigorously, including higher orders in perturbation theory, where non-Abelian interactions become relevant.

In this paper, we have computed $\order{\mg}$ corrections to the differential $t \bar t$ production cross section through NNLO.
We have found that such corrections cancel in the sum of many different contributions, provided that the cross section is expressed in terms of a short-distance mass of the top quark.
Since this statement generalizes the known
one-loop result \cite{Makarov:2023uet}, it is perhaps not very surprising.
However, it strongly suggests the existence of a universal mechanism underlying this cancellation, whose understanding may lead to an extension of the KLN theorem to linear power corrections at all perturbative orders.

Our computation employs advanced methods of multi-loop calculations such as asymptotic expansion of Feynman diagrams, integration-by-parts techniques augmented with reverse unitarity, and momentum remappings in the real-emission contributions.
These methods allow us to demonstrate the cancellation of $\order{\mg}$ contributions to the $t \bar t$ production cross section at NLO in a straightforward way.
However, a similar analysis at NNLO turns out to be considerably more involved, even though advanced technical methods are used.

In this respect, it is interesting to point out an unusual aspect of our calculation, i.e.\ the fact that, in order to establish the cancellation of the $\order{\mg}$ terms, we \emph{do not need to calculate} the master integrals in the usual sense of this word. 
Rather, by integrating over energies of virtual gluons, which can be done using 
Cauchy's theorem, we demonstrate that all two-loop integrals that contribute to $\order{\mg}$ corrections to the $t \bar t$ production cross section at $\order{\as^2}$ can be written as linear combinations of  real-virtual and double-real integrals.  Once this is done, the cancellation of ${\cal O}(m_g)$ terms happens without the need to compute the latter. 

The established cancellation of $\order{\mg}$ contributions to the $t \bar t$ production cross section -- together with the manner in which virtual and real-emission contributions, as well as renormalization constants, cancel among themselves -- suggests that there should be a more direct and physical way to arrive at this result.
To derive it, one can extend this study  in several directions.
First, it would be very interesting to understand how this calculation can be performed without using integration-by-parts identities.
Doing this would require establishing relations between next-to-soft corrections to loop and real-emission amplitudes that describe the $t \bar t$ production process, along the lines of what has been done in Refs~\cite{Makarov:2023uet,Makarov:2023ttq, Czakon:2025cxk}. 
Such relations
will be critical for proving the cancellation of linear power corrections to more complex processes at high perturbative orders where  exact calculations become impossible.
Second, it would be important to check whether a similar mechanism applies to collider processes with massless energetic particles both in the final and initial states.
The latter case is especially important, since its understanding is essential for claiming  such cancellations for processes at hadron colliders.
Finally, one should
develop a framework that connects these results to actual QCD.
This may require reformulating the above findings in
an operator-based language.

%-----------------------------------------------------
%  ACKNOWLEDGMENTS
%-----------------------------------------------------

\acknowledgments
This research was supported
by the Deutsche Forschungsgemeinschaft (DFG, German Research Foundation) under grant 396021762 - TRR 257. 
D.F. thanks the Institute for Theoretical Physics at the University of Bern for hospitality and support.
D.H. and K.M. thank the Erwin-Schrödinger International Institute for Mathematics and Physics at the University of Vienna for partial support during the Programme `New Paradigms for Harnessing Quantum Field Theory at Colliders', July 27 – August 28, 2026.
All Feynman diagrams were drawn using \texttt{TikZ-Feynman}~\cite{Ellis:2016jkw}.

%%%%%%%%%%%%%%%%%%%%%%%%%%%%%%%%%%%%%
%%% APPENDIX 
%%%%%%%%%%%%%%%%%%%%%%%%%%%%%%%%%%%%%
%\newpage
\appendix

\section{Momentum redefinitions for  real-emission contributions}
\label{appendix:kSqRealToBorn}

The ease and efficiency with which the cancellation of 
${\cal O}(m_g)$ contributions   
to physical processes can be investigated rely on the 
possibility of factorizing the 
real-emission phase spaces for 
one- and two-gluon emissions with 
${\cal O}(m_g)$ accuracy. 
This factorization is enabled by momentum transformations that were  discussed 
in Refs~\cite{Makarov:2023ttq,Makarov:2023uet}.  
The goal of this Appendix is to 
generalize the discussion provided 
in those 
references. 

To this end, we consider the 
following transformation 
\begin{equation} \label{eq:RealToBornDefinitionCopy}
\begin{aligned}
    \qt &= \pt - \frac{1+\rbchi}{2} \, k + A\,\pt + B\,\ptb \, , \\[.2em]
    \qtb &= \ptb - \frac{1-\rbchi}{2} \, k - A\,\pt - B\, \ptb \, ,
\end{aligned}    
\end{equation}
where $k$ can be either the four-momentum 
of a single massive gluon or 
the sum of the four-momenta of two massive gluons, $\chi$ is an 
arbitrary parameter, and 
$A,B$ are functions of $\pt,\ptb$ and $k$.
We assume that $k \sim  m_g$ 
and $k^2 \sim m_g^2$.
We can choose functions $A,B$ in such a way that  the conditions 
$\qt^2= \pt^2$ and $\qtb^2 = p_{\bar t}^2 $ are satisfied to the required order in the  
$\order{\mg}$ expansion.

Since 
\begin{equation}
    \qt + \qtb + k  = \pt + \ptb \,, 
\end{equation}
the gluon momentum is removed  from the energy-momentum-conserving $\delta$-function.
Then,  to factorize the phase space, we write  
\begin{equation}
    \rmd^d \qt \, \rmd^d \qtb 
    = \rmd^d \pt \, \rmd^d \ptb \, 
    J \,,
    \label{eqA.3}
\end{equation}
where $J$ is the Jacobian of the transformation. Our goal is to determine it. 

As the first step, we perform an ancillary transformation 
\begin{equation}
    q_\Sigma = \qt + \qtb \, , 
    \qquad  r = \qt - \qtb \, ,
    \end{equation}
    and obtain 
    \begin{equation}
    \rmd^d \qt \, \rmd^d \qtb = N \, \rmd^d q_\Sigma \, \rmd^d r \, , 
\end{equation}
where $N$ is a momentum-independent 
factor. Introducing 
\begin{equation}
    P = \pt + \ptb \,, 
    \qquad 
    \pi = \pt - \ptb \,, 
\end{equation}
we find 
\begin{equation}
    q_\Sigma= P - k \,, 
    \qquad  
    r = \pi - \rbchi\, k 
    + (A+B) P +(A-B) \pi \, .
\end{equation}
It follows that  
\begin{equation}
    \rmd^d \qt \, \rmd^d \qtb 
    = N \rmd^d P \, \rmd^d \pi \left | \frac{\partial r^\mu}{\partial \pi^\nu } \right | 
    = \rmd^d \pt \, \rmd^d \ptb 
    \left | \frac{\partial r^\mu }{\partial \pi^\nu} \right |. 
\end{equation}
This equation implies that the required Jacobian reads 
\begin{equation}
J = \left | \frac{\partial r^\mu }{\partial \pi^\nu} \right |,
\end{equation}
where the derivative has to be computed at fixed $P$ and $k$.
We find 
\begin{equation}
R^{\mu \nu} = \frac{\partial r^\mu }{\partial \pi_\nu}
= g^{\mu \nu} (1 + A - B) 
+P^\mu U^\nu 
+ \pi^\mu V^\nu, 
\end{equation}
where 
\begin{equation}
    U^\nu = \frac{\partial (A+B)}{
    \partial \pi_\nu
    } \,,
    \qquad
    V^\nu = \frac{\partial (A-B)}{
    \partial \pi_\nu} \, .
\end{equation}
To find the determinant 
of the matrix $R^{\mu \nu}$, we compute
its eigenvalues.  To this end, we consider 
the $d$-dimensional vector space 
and choose $d-2$ vectors that are orthogonal to the vectors 
$U$ and 
$V$.  If $\xi$ is such a vector, then 
\begin{equation}
R^{\mu \nu} \xi_\nu = \Delta \; \xi^\mu \, ,
\end{equation}
with $\Delta = 1+A-B$.
To find the remaining two vectors, we consider a vector that is a linear combination of 
$P^\mu$ and $\pi^\nu$,
i.e. 
\begin{equation}
    \xi^\mu = \alpha P^\mu + \beta \pi^\mu \, . 
\end{equation}
Then, by requiring 
\begin{equation}
R^{\mu \nu} \xi_\nu = \lambda \;  \xi^\mu \, , 
\end{equation}
we conclude  that finding the two missing eigenvalues is equivalent to computing the determinant of the following two-by-two matrix 
\begin{equation}
{\rm det} 
\left (
\begin{array}{cc} 
\Delta + P \cdot U &  \pi \cdot U
\\
P \cdot V &  \Delta + \pi \cdot V
\end{array}
\right )
= {\rm det} 
\left ( 
\Delta \mathbb{1}_2 + {\cal D}
\right ), 
\end{equation}
where $\mathbb{1}_2$ is the unity matrix and ${\cal D}$ 
reads 
\begin{align}
    {\cal D} =
\begin{pmatrix}
R_t A&R_{\bar t}A\\
R_t B&R_{\bar t}B
\end{pmatrix}. 
\end{align}
In the above equation $R_t$ and $R_{\bar t}$ are directional derivatives defined by
\begin{equation} \label{eq:DefDirectedDerivatives}
R_t f = \pt^\mu\left( \frac{\partial f}{\partial \pt^\mu} - \frac{\partial f}{\partial  \ptb^\mu} \right) \,,
\qquad
R_{\bar t} f = \ptb^\mu\left( \frac{\partial f}{\partial \pt^\mu} - \frac{\partial f}{\partial \ptb^\mu} \right) \, , 
\end{equation}
for a test function $f$. 
Hence, we conclude that  the Jacobian
in Eq.~(\ref{eqA.3}) evaluates to 
\begin{equation}
J = \Delta^{d-2} {\rm det} 
\left ( 
\Delta \mathbb{1}_2 + {\cal D}
\right ) \, .
\end{equation}
Using this result, we obtain 
the desired phase-space transformation 
\begin{align}
     {\rm dLips}(P; \qt,\qtb, k)
    = {\rm dLips_{LO}}(P; \pt,\ptb) \, \frac{{\rm d}^d k}{(2 \pi)^{d-1}}  \delta_+(k^2 - \mg^2) \Delta^{d-2} \det(\Delta \mathbb{1}_2 + {\cal D}) 
       \, , 
\end{align}
where the mass-shell limit for momenta 
$p_{t, \bar t}$ 
needs to be taken \emph{after} 
the derivatives in ${\cal D}$ are evaluated.  We emphasize that the above formula should be understood as the generating function 
for expanding the 
$\bar t t g$ phase space at small values of the gluon momentum $k \sim m_g$, rather than the exact transformation valid for arbitrary $k$. 

The last required  ingredients are 
the functions $A,B$. To 
${\cal O}(k^2)$ accuracy, they are easily found to be
\begin{align}
    A =&-\frac{
    (1+\rbchi)\,\ptb^2\,(\pt \cdot k)
    +(1-\rbchi)\,(\pt \cdot \ptb)\,(\ptb \cdot k)
    }{
    2\left[(\pt \cdot \ptb)^2-\pt^2\ptb^2\right]
    } \\
    &+\frac{1}{
    8\left[(\pt \cdot \ptb)^2-\pt^2\ptb^2\right]^2
    }
    \bigg\{
    (1-\rbchi)\,\pt^2\,(\ptb \cdot k)^2
    \Big[
    (1-\rbchi)(\pt \cdot \ptb)
    -(3+\rbchi)\ptb^2
    \Big] \nonumber \\
    &\quad -2\,(\pt \cdot k)(\ptb \cdot k)(\pt \cdot \ptb)
    \Big[
    (1+\rbchi)^2\ptb^2
    +(1-\rbchi)^2(\pt \cdot \ptb)
    \Big] \nonumber \\
    &\quad
    +(1+\rbchi)\,\ptb^2\,(\pt \cdot k)^2
    \Big[
    (1+\rbchi)\ptb^2
    -(3-\rbchi)(\pt \cdot \ptb)
    \Big] \nonumber \\
    &\quad
    +k^2
    \Big[
    (1+\rbchi)^2\ptb^2
    +(1-\rbchi)^2(\pt \cdot \ptb)
    \Big]
    \big[
    (\pt \cdot \ptb)^2-\pt^2\ptb^2
    \big]
    \bigg\}
    + \order{k^3} \,, \notag \\
    B = & -A(\rbchi \to -\rbchi,
         \pt \leftrightarrow \ptb)\, .
\end{align}
\section{List of master integrals}
\label{appendix_sec_list_MIs}
In this appendix, we collect all master integrals required for the calculation of the NLO and NNLO corrections described in the main text.
We denote the regular and eikonal (massive-fermion) propagators as
\begin{equation}
    G_x(k) = \frac{1}{k^2 - m_x^2 + \iZero} \,,
    \qquad
    F_p(k) = \frac{1}{2 k\cdot p + \iZero} \,,
\end{equation}
where $p$ is an external four-momentum vector, which can be either $\Fm$ or $\Sm$ introduced in \eq\eqref{eq_amu_bmu_def}, while the subscript $x$ in $G_x(k)$ identifies the mass of the particle associated with that propagator.
The soft master integrals listed in Table~\ref{tab_master_integrals} contain $G_g$ for gluons, $G_0$ for massless quarks, $G_t$ for the top quark, and $G_H$ for the Higgs boson.
For simplicity, we omit the subscript $g$ from $G_g$ in what follows.
The notations $\langle\mydots\rangle$ and $\langle\mydots\rangle_i$ are defined in Eqs \eqref{eq6.5} and \eqref{eq:6.9}, respectively.
The momentum $h$ appearing in $\IV{4}$ satisfies $h^2 = \mg^2$.
{
\renewcommand{\arraystretch}{1.35}
\begin{longtable}{
  @{} >{$}l<{$} @{}
      >{$}l<{$} @{}
      >{$}l<{$} @{\quad}
      >{$}l<{$} @{}
}
\caption{List of master integrals.}
\label{tab_master_integrals} \\
%----------------------------------------------
% IVV
%----------------------------------------------
\multicolumn{4}{c}{\underline{Double-virtual master integrals}} \\
    & \IVV{1} = \autoL G(k_1) G(k_2) G(k_{12}^-) F_\Fm(k_1) F_\Sm(-k_1) F_\Sm(-k_2) \autoR 
    & & \IVV{6} = \autoL G(k_1) G(k_2) F_\Fm(k_1) F_\Sm^2(-k_{12}^+) \autoR \\
    & \IVV{2} = \autoL G(k_1) G(k_2) F_\Fm(k_1) F_\Sm(-k_1) F_\Sm(-k_2) \autoR 
    & & \IVV{7} = \autoL G(k_1) G_0(k_2) F_\Fm(k_1) G_0(k_{12}^+) \autoR \\
    & \IVV{3} = \autoL G(k_1) G(k_2) F_\Fm(k_1) F_\Sm(-k_1) F_\Fm(k_{12}^+) \autoR 
    & & \IVV{8} = \autoL G(k_1) G(k_2) G(k_{12}^+) F_\Fm(k_1) \autoR \\
    & \IVV{4} = \autoL G(k_1) G(k_2) F_\Sm(-k_1) F_\Fm(k_{12}^+) F_\Sm(-k_{12}^+) \autoR 
    & & \IVV{9} = \autoL G(k_1) G(k_2) F_\Sm(-k_{12}^+) \autoR \\
    & \IVV{5} = \autoL G(k_1) G(k_2) G(k_{12}^-) F_\Fm^2(k_1) F_\Sm(-k_2) \autoR 
    & & \IVV{10} = \autoL G(k_1) G(k_2) F_\Fm(k_1) \autoR \\[7pt]
%----------------------------------------------
% IRV
%----------------------------------------------
\multicolumn{4}{c}{\underline{Real-virtual master integrals}} \\
    & \IRV{1} = \autoL G(k_2) G(k_{12}^+) F_\Fm(k_1) F_\Sm(k_1) F_\Sm(-k_2) \autoR_1 
    & & \IRV{10} = \autoL G(k_2) F_\Fm(k_1) F_\Sm(k_1) F_\Sm(-k_2) \autoR_1 \\
    & \IRV{2} = \autoL G(k_2) G(k_{12}^-) F_\Sm(k_1) F_\Fm(k_2) F_\Sm(k_{12}^-) \autoR_1 
    & & \IRV{11} = \autoL G(k_2) G(k_{12}^-) F_\Fm(k_2) \autoR_1 \\
    & \IRV{3} = \autoL G(k_2) F_\Fm(k_1) F_\Sm(k_1) F_\Fm(k_{12}^+) \autoR_1 
    & & \IRV{12} = \autoL G(k_2) G(k_{12}^-) F_\Sm(k_1) \autoR_1 \\
    & \IRV{4} = \autoL G(k_2) G(k_{12}^-) F_\Fm^2(k_2) F_\Sm(k_{12}^-) \autoR_1 
    & & \IRV{13} = \autoL G(k_2) F_\Fm(k_{12}^+) F_\Sm^2(-k_2) \autoR_1 \\
    & \IRV{5} = \autoL G(k_2) G(k_{12}^-) F_\Sm(k_1) F_\Fm^2(k_2) \autoR_1 
    & & \IRV{14} = \autoL G(k_2) F_\Sm(k_1) F_\Fm^2(k_{12}^+) \autoR_1 \\
    & \IRV{6} = \autoL G(k_2) G(k_{12}^-) F_\Sm^2(k_1) F_\Fm(k_2) \autoR_1 
    & & \IRV{15} = \autoL G(k_2) F_\Fm(k_{12}^+) \autoR_1 \\
    & \IRV{7} = \autoL G(k_2) F_\Fm(k_2) F_\Fm(k_{12}^+) F_\Sm(-k_2) \autoR_1 
    & & \IRV{16} = \autoL G(k_2) F_\Fm(k_2) \autoR_1 \\
    & \IRV{8} = \autoL G(k_2) F_\Sm(k_1) F_\Fm(k_{12}^+) F_\Sm(-k_2) \autoR_1 
    & & \IRV{17} = \autoL G(k_2) F_\Sm(k_1) \autoR_1 \\
    & \IRV{9} = \autoL G(k_2) F_\Sm(k_1) F_\Fm(k_2) F_\Sm(-k_2) \autoR_1 
    & & \\[7pt]
%----------------------------------------------
% IRR
%----------------------------------------------
\multicolumn{4}{c}{\underline{Double-real master integrals}} \\
    & \IRR{1} = \autoL G(k_{12}^+) F_\Sm(k_1) F_\Fm(k_2) F_\Sm(k_{12}^+) \autoR_{12} 
    & & \IRR{8} = \autoL F_\Fm(k_2) F_\Fm(k_{12}^+) F_\Sm(k_2) \autoR_{12} \\
    & \IRR{2} = \autoL G(k_{12}^+) F_\Fm(k_{12}^+) F_\Sm(k_2) F_\Sm(k_{12}^+) \autoR_{12} 
    & & \IRR{9} = \autoL G(k_{12}^+) F_\Sm(k_2) \autoR_{12} \\
    & \IRR{3} = \autoL G(k_{12}^+) F_\Sm^2(k_1) F_\Fm(k_2) \autoR_{12} 
    & & \IRR{10} = \autoL G(k_{12}^+) F_\Fm(k_{12}^+) \autoR_{12} \\
    & \IRR{4} = \autoL F_\Fm(k_1) F_\Sm(k_1) F_\Fm(k_2) \autoR_{12} 
    & & \IRR{11} = \autoL F_\Fm(k_{12}^+) F_\Sm^2(k_2) \autoR_{12} \\
    & \IRR{5} = \autoL G(k_{12}^+) F_\Fm(k_{12}^+) F_\Sm^2(k_2) \autoR_{12} 
    & & \IRR{12} = \autoL F_\Fm(k_{12}^+) \autoR_{12} \\
    & \IRR{6} = \autoL G(k_{12}^+) F_\Fm^2(k_{12}^+) F_\Sm(k_2) \autoR_{12} 
    & & \IRR{13} = \autoL F_\Fm(k_2) \autoR_{12} \\
    & \IRR{7} = \autoL F_\Fm(k_2) F_\Fm(k_{12}^+) F_\Sm(k_{12}^+) \autoR_{12} 
    & & \IRR{14} = \Re\autoL G(k_{12}^+) F_\Fm(k_{12}^+) \autoR_{1_0 2_0} \\[7pt]
%----------------------------------------------
% IV
%----------------------------------------------
\multicolumn{4}{c}{\underline{Single-virtual master integrals}} \\
    & \IV{1} = \autoL G(k_1) F_\Fm(k_1) \autoR
    & & \IV{5} = \autoL G(k_1) G(k_1-h) \autoR
    \\
    & \IV{2} = \autoL G_t(k_1) \autoR
    & & \IV{6} = \autoL G_H(k_1) \autoR
    \\
    & \IV{3} = \autoL G(k_1) \autoR
    & & \IV{7} = \autoL G(k_1) F_\Fm(k_1) F_\Sm(-k_1) \autoR
    \\
    & \IV{4} = \autoL G_0(k_1) G_0(k_1-h) \autoR
    & & \IV{8} = \autoL G_t(k_1) G_t(k_1-\pt-\ptb) \autoR \\[7pt]
%----------------------------------------------
% IR
%----------------------------------------------
\multicolumn{4}{c}{\underline{Single-real master integrals}} \\
    & \IR{1} = \autoL 1 \autoR_1
    & & \IR{3} = \autoL F_\Fm(k_1) F_\Sm(k_1) \autoR_1
    \\
    & \IR{2} = \autoL F_\Fm(k_1) \autoR_1
    & &
\end{longtable}
}

%%%%%%%%%%%%%%%%%%%%%%%%%%%%%%%%%%%%%
%%% BIBLIOGRAPHY 
%%%%%%%%%%%%%%%%%%%%%%%%%%%%%%%%%%%%%

%\newpage
\bibliographystyle{JHEP}
\bibliography{biblio.bib}

\end{document}